\documentclass[journal]{IEEEtran}
\usepackage{amsfonts}
\usepackage{booktabs}
\usepackage{cite}

\usepackage{makecell,multirow,diagbox}
\usepackage{pifont}
\usepackage{enumerate}

\ifCLASSINFOpdf
   \usepackage[pdftex]{graphicx}
   \graphicspath{{../pdf/}{../jpeg/}}
   \DeclareGraphicsExtensions{.pdf,.jpeg,.png}
\else
   \usepackage[dvips]{graphicx}
   \graphicspath{{../eps/}}
   \DeclareGraphicsExtensions{.eps}
\fi
\usepackage[cmex10]{amsmath}

\usepackage{mathabx}
\usepackage{algorithm}
\usepackage{algorithmic}
\usepackage{color, soul}
\soulregister\cite7 
\soulregister\citep7 
\soulregister\citet7 
\soulregister\ref7 
\soulregister\eqref7 
\soulregister\pageref7 
\definecolor{shadecolor}{rgb}{0.92,0.92,0.92}
\usepackage{framed}
\begin{document}
%
\title{Mobile Device Association and Resource Allocation in Small-Cell IoT Networks with Mobile Edge Computing and Caching}


\author{Tianqing~Zhou,~
        Yali~Yue,~
        Dong~Qin,~
        Xuefang~Nie,~
        Xuan~Li,~
        and Chunguo~Li
\thanks{This work was supported by National Natural Science Foundation of China under Grant Nos. 61861017, 61671144, 61861018, 61761030, 61961020, 61862025, 61663010 and 61963017, Foundation of Jiangxi Educational Committee of China under Grant Nos. GJJ180312 and GJJ180313.}
\thanks{T. Zhou, Y. Yue, X. Nie and X. Li are with the School of Information Engineering, East China Jiaotong University, Nanchang 330013, China (email: zhoutian930@163.com; yueyali1015@163.com; Xuefangnie@163.com; lixuan0799@outlook.com).}
\thanks{D. Qin is with School of Information Engineering, Nanchang University, Nanchang 330031, China (e-mail: qindong@seu.edu.cn).}
\thanks{C. Li is with School of Information Science and Engineering, Southeast University, Nanjing 210096, China (email: chunguoli@seu.edu.cn)}
}


%


\maketitle

\begin{abstract}
To meet the need of computation-sensitive (CS) and high-rate (HR) communications, the framework of mobile edge computing and caching has been widely regarded as a promising solution. When such a framework is implemented in small-cell IoT (Internet of Tings) networks, it is a key and open topic how to assign mobile edge computing and caching servers to mobile devices (MDs) with CS and HR communications. Since these servers are integrated into small base stations (BSs), the assignment of them refers to not only the BS selection (i.e., MD association), but also the selection of computing and caching modes. To mitigate the network interference and thus enhance the system performance, some highly-effective resource partitioning mechanisms are introduced for access and backhaul links firstly. After that a problem with minimizing the sum of MDs' weighted delays is formulated to attain a goal of joint MD association and resource allocation under limited resources. Considering that the MD association and resource allocation parameters are coupling in such a formulated problem, we develop an alternating optimization algorithm according to the coalitional game and convex optimization theorems. To ensure that the designed algorithm begins from a feasible initial solution, we develop an initiation algorithm according to the conventional best channel association, which is used for comparison and the input of coalition game in the simulation. Simulation results show that the algorithm designed for minimizing the sum of MDs' weighted delays may achieve a better performance than the initiation (best channel association) algorithm in general.
\end{abstract}
\begin{IEEEkeywords}
Mobile device association, IoT networks, mobile edge computing, mobile edge caching, resource allocation, delay minimization.
\end{IEEEkeywords}

%
\IEEEpeerreviewmaketitle

\section{Introduction}
In recent years, more and more new computation-sensitive (CS) and high-rate (HR) services (e.g., mobile gaming, cognitive assistance and virtual/augmented reality, etc.) are driven by mobile computing applications and Internet of Things (IoT). Although the cloud computing is widely advocated for highly-effective resource utilization \cite{TTaleb2017,YMao2017,SWang2017,JYao2019}, it may be unable to meet some stringent requirements of these latency-sensitive services. In addition, it may also be impractical to transport all data traffic to the remote cloud over today's highly-congested backbone networks, especially for staggeringly-growing distributed data traffic. In fact, the essential challenges caused by the popularity of these services may be how to deal with the shortage of spectral and backhaul resources, the limitation of computing capability and energy at mobile terminals, the high quality-of-experience (QoE) requirements of services, etc.
\par
To overcome those aforementioned challenges, the mobile edge computing and caching techniques have been recently regarded as important and indispensable parts of new network paradigm \cite{PMach2017,FGuo2018Dec,LChen2018Aug,JZhao2019Aug,KPoularakis2016Apr,LWang2018Sep,NAbbas2018Sep}, which push the frontier of data services and computing applications away from the cloud computing center to the logical edge of networks, thereby enabling the knowledge generation and analytic to occur in close proximity to data sources. By deploying mobile edge computing and caching servers in small-cell (IoT) networks \cite{Zhou2018Mar,TZhou2019May,JZhao2019Dec}, mobile base stations (BSs) are endowed with cloud-like storage and computing capability, and thus may be seen as a substitute of cloud computing.
\par
In small-cell IoT networks with mobile edge computing and caching, by implementing the caching servers to small BSs (SBSs), the caching technology changes the traditional way of reactive or even passive content (data) request and transmission, and thus it can be used for supporting the high QoE requirements of services and ensuring HR communications. In fact, the application of caching servers can not only shorten the distance between mobile terminals (destinations) and data sources, but it can also greatly reduce the resource consumption of backhaul links. Certainly, in order to further improve the data access efficiency, the caching servers can proactively predict some popular contents and then cache them. In addition, by implementing computing servers to SBSs, the edge computing technology enables cloud-computing capabilities to be closer to mobile terminals, and thus some challenges including limited battery and computation capacity can be overcome well.
\par
Although the deployment of computing and caching servers can greatly support CS and HR services and thus guarantees high QoE requirements, it may make the mobile device (MD) association more complicated. The MD association often selects some nodes (BSs) for MDs to optimize some certain performance metric under some necessary constraints \cite{QYe2016May,DLiu2016Sec,YXu2017Mar}. In small-cell IoT networks with mobile edge computing and caching severs, the MD association refers to not only the BS selection, but also the selection of computing and caching modes. That means the MD association may be complicated and challenging in such a new framework. Significantly, the computing modes refer to edge and local computing functions, and the caching modes include the direct transfer that retrieves files from SBSs directly and the indirect one that gets files from SBSs through backhaul links between them and macro BSs (MBSs).

\subsection{Related Work}
In small-cell IoT networks, the existing works, closely related to mobile edge computing or/and caching, can be roughly divided into the following three groups.
\par
In the first group, the MD association is just tightly related to mobile edge caching. In \cite{GRen2017Oct}, a distributed MD association was performed to achieve a balance between rate utility and caching data rate for cache-enabled small-cell networks. In \cite{THan2017Oct}, a MD association mechanism was designed to achieve a tradeoff between green communications and network delay for cache-enabled small-cell networks with limited backhaul capacity. Similarly, another MD association in \cite{JKwak2019Aug} was developed to achieve such a tradeoff for multi-cell networks under two time-scales. Unlike the hybrid power supply, designers in \cite{FGuo2018Jul} introduced energy harvesting techniques, and designed a MD association mechanism to maximize the number of requests handled at BSs for small-cell networks. In addition, designers in \cite{KGuo2018Nov} jointly optimized the MD association and BS muting to maximize the number of served MDs. Besides these mechanisms, there still exist other association mechanisms that jointly executed the MD association and content caching to minimize distinct network delay \cite{WTeng2019Oct,YWang2016}, and another delay-aware association mechanism \cite{WJing2019}.
\par
In the second group, the MD association just has a close relation with mobile edge computing. In \cite{YDai2018Dec}, the joint optimization of MD association and computation offloading was performed to achieve a goal of energy consumption minimization for a mobile edge computing system with multiple tasks. In \cite{YDu2019}, a two-tier matching algorithm was developed to solve a problem of computing resource management and trading for small-cell networks. In \cite{YQian2019Oct}, the joint optimization of uploading power of MDs, MD association and UAV (unmanned aerial vehicle) trajectory was performed to maximize the sum bit offloaded from MDs to UAVs. In fact, among this type of studies, it is easy to find that there exist very few ones concentrated on the network delay.
\par
In the last group, the MD association has a tight relation with both mobile edge computing and caching. So far, such a type of studies is relatively few and is still an open topic. In \cite{JZhou2019}, the MD association was performed to minimize weighted network delay under limited computation and storage capabilities, but didn't mention the resource allocation. In \cite{ZTan2018Feb}, the designers tried to perform the MD association and resource allocation to optimize network throughput, computing and caching cost.
\par
In general, the first two groups just concentrate on the one of mobile edge computing and caching, and most of association mechanisms didn't mention the resource allocation in the last group.

\subsection{Contributions and Organization}
 In this paper, we perform the MD association for small-cell IoT networks with mobile edge computing and caching, which differs from the first two groups of existing works. Specifically, we jointly consider the mobile edge computing and caching during the MD association, but the latter just mentions the one of them. Unlike the third type of existing works, we jointly perform the MD association and resource allocation to minimize the sum of MDs' weighted delays under a distinct network framework. In detail, the main contributions of this paper can be listed as follows.
 \par
 \textbf{\ding{182} Association Design under Orthogonal Resource Partitioning.} Instead of full duplex used for interference mitigation in the existing related works, we consider the orthogonal time partitioning for downlink and uplink access links, and the orthogonal frequency partitioning for backhaul and access links. So far, such a work should be a completely new investigation for joint MD association and resource allocation in small-cell IoT networks with mobile edge computing and caching.
 \par
 \textbf{\ding{183} New Association Mechanism Development.} Under the orthogonal resource partitioning, we jointly consider the MD association and resource allocation for small-cell IoT networks with mobile edge computing and caching, minimizing the sum of MDs' weighted delays under the constraints of limited resources. In addition, according to conventional channel-gain-based association, we also design the best channel association mechanism with equal resource allocation under resource constraints, which is also regarded as an important part of the former association mechanism.
 \par
 \textbf{\ding{184} The Design of Association Algorithm.} To solve the formulated problem with minimizing the sum of MDs' weighted delays, we develop an alternating optimization algorithm by decoupling MD association and resource allocation parameters, where the coalitional game and convex optimization theorems are used for solving the MD association and resource allocation subproblems respectively. In addition, we also develop an algorithm for best channel association mechanism, which is used for the initiation of alternating optimization algorithm and comparison in the simulation.
 \par
\textbf{\ding{185} Necessary Analyses of Designed Algorithms.} At first, we provide some convergence proofs for the MD association and resource allocation subalgorithms. Secondly, the stability analysis of coalitional game is shown. At last, some detailed computation complexity analyses are given for all designed algorithms, and the corresponding parallel implement of them is also discussed in detail.
\par
The rest of this paper is organized as follows. In Section II, the network, resource, communication, computation and cache models are illustrated. In Section III, a problem with minimizing the sum of MDs' weighted delays, jointly investigating the MD association and resource allocation, is formulated. In Section IV, two types of algorithms are designed for joint MD association and resource allocation. In Section V, the detailed analyses of stability, convergence, computation complexity and parallel implement of designed algorithms are provided. In Section VI, the performance evaluation is performed. In Section VII, some conclusions and further discussions are made.

\section{SYSTEM MODEL}
\subsection{Network Model}
\begin{figure}[!t]
\centering
\centerline{\includegraphics[width=3.5in,height=2.5in]{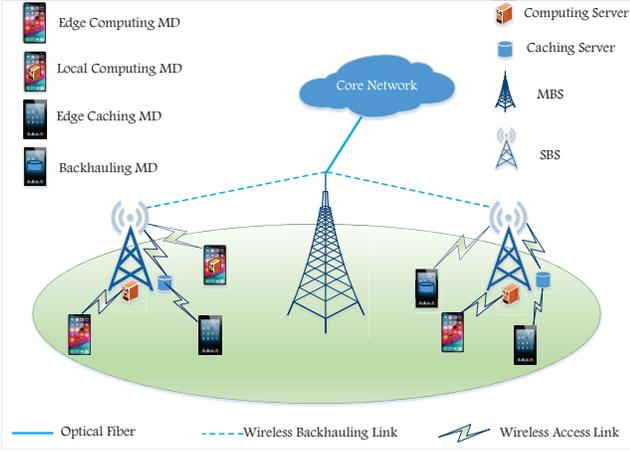}}
\caption{Network Deployment.}
\label{fig1}
\end{figure}

In this paper, we concentrate on small-cell IoT networks with mobile edge computing and caching, which are illustrated in Fig.\ref{fig1}. There exist ${\Hat{N}}$ MBSs in ${\Hat{\mathcal{N}}}=\big\{ 1,2,\cdots ,{\Hat{N}} \big\}$, and ${\Bar{N}}$ SBSs in ${\Bar{\mathcal{N}}}=\left\{ 1,2,\cdots ,{\Bar{N}} \right\}$. In addition, ${\Bar{K}}$ HR MDs (HRDs) are in ${\Bar{\mathcal{K}}}=\left\{ 1,2,\cdots ,{\Bar{K}} \right\}$, which include edge caching ones and backhauling ones. ${\Hat{K}}$ CS MDs (CSDs) are in ${\Hat{\mathcal{K}}}=\big\{ 1,2,\cdots ,{\Hat{K}} \big\}$, which include edge computing ones and local computing ones. It is noteworthy that the only wireless backhaul links are supported between SBSs and MBSs. At each SBS, one mobile edge computing server and one mobile edge caching server are implemented. When a required content has been cached at one SBS selected by some edge caching MD, it can be retrieved by this MD directly. If the required content has not been cached at any SBS, a backhauling MD can just retrieve it from core network through some backhaul link between SBS and MBS. In our model, any computation task of local computing MD can be executed at mobile terminal, and the one of edge computing MD can be offloaded to a mobile edge computing server.
\par
As shown in Fig.\ref{fig1}, each SBS just implements only one computing server and only one caching server. That means the selection of mobile edge computing or caching server refers to the one of SBS in fact.
\subsection{Resource Model}
\begin{figure}[!t]
\centering
\centerline{\includegraphics[width=3in,height=1.5in]{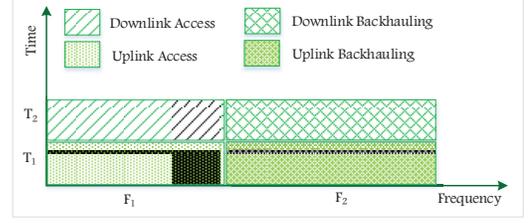}}
\caption{Resource Partitioning.}
\label{fig2}
\end{figure}

To mitigate the network interference, the orthogonal frequency and time partitioning scheme is utilized, and shown in Fig.\ref{fig2}. In it, the width of frequency band ${F}_{1}$ is $a{W}$, the one of frequency band ${F}_{2}$ is equal to $\left( 1-a \right){W}$, where $a\in \left[ 0,1 \right]$ and ${W}$ represent the frequency partitioning factor and system bandwidth respectively; the channel coherence block ${T}$ contains two time slots (i.e., ${T}_{1}$ and ${T}_{2}$) satisfying ${{{T}}_{1}}+{{{T}}_{2}}={T}$, where ${T}_{1}$ and ${T}_{2}$ represent the time slots used for uplink and downlink transmission in a channel coherence block respectively. To proceed, some necessary assumptions on the resource utilization are listed as follows.
\begin{figure}[!t]
\centering
\centerline{\includegraphics[width=3in,height=2.5in]{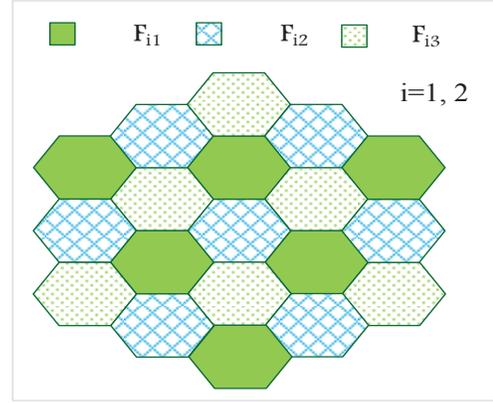}}
\caption{Frequency Utilization in the Uplink or Downlink.}
\label{fig3}
\end{figure}
\par
\noindent
\textit{\textbf{Assumption 1: }}MDs utilize the frequency band ${F}_{1}$ for uplink and downlink access links, but adopt the time slots ${T}_{1}$ and ${T}_{2}$ for them respectively.
\par
\noindent
\textit{\textbf{Assumption 2: }}MDs use the frequency band ${F}_{2}$ for uplink and downlink backhaul links, but employ the time slots ${T}_{1}$ and ${T}_{2}$ for them respectively.
\par
\noindent
\textit{\textbf{Assumption 3: }}The three subbands (i.e., ${F}_{11}$, ${F}_{12}$ and ${F}_{13}$) of ${F}_{1}$ are utilized by different adjacent clusters on both uplink and downlink access links, where ${F}_{11}$, ${F}_{12}$ and ${F}_{13}$ have equal bandwidth; each macrocell represents a cluster for simplicity.
\par
\noindent
\textit{\textbf{Assumption 4: }}The three subbands (i.e., ${F}_{21}$, ${F}_{22}$ and ${F}_{23}$) of ${F}_{2}$ are utilized by different adjacent clusters on both uplink and downlink backhaul links, where ${F}_{21}$, ${F}_{22}$ and ${F}_{23}$ have equal bandwidth.
\par
By employing different time slots, the interference between uplink and downlink access links may be cancelled, and the one between uplink and downlink backhual links may also be cancelled. By implementing distinct frequency bands, the interference between access and backhaul links may be cancelled. In addition, Assumption 3 shows that the inter-cell interference between any two macrocells may be cancelled by employing distinct subbands for them on both uplink and downlink access links. Similarly, Assumption 4 shows that such a type of interference may be cancelled on both uplink and downlink backhaul links. The detailed illustration of Assumption 3 and Assumption 4 can be found in Fig.\ref{fig3}.

\subsection{Communication Model}
In the communication model, we make some following assumptions on the content retrieve and resource usage.
\par
\noindent
\textit{\textbf{Assumption 5: }}MDs may request some files from $\mathcal{I}=\left\{ 1,2,\cdots ,I \right\}$, and each file request is entirely served by one SBS. In addition, all files in $\mathcal{I}$ have the same size, which means the file size of any file $i$ satisfying ${{\Bar{d}}_{i}}=L$.
\par
\noindent
\textit{\textbf{Assumption 6: }}Any subbdand (e.g., ${F}_{11}$, ${F}_{12}$ or ${F}_{13}$) is equally allocated to all SBSs at each macrocell on both uplink and downlink access links.
\par
\noindent
\textit{\textbf{Assumption 7: }}Any subbdand (e.g., ${F}_{21}$, ${F}_{22}$ or ${F}_{23}$) is equally allocated to all SBSs at each macrocell on both uplink and downlink backhaul links.
\par
\noindent
\textit{\textbf{Assumption 8: }}Different subband slices of some subband (e.g., ${F}_{11}$, ${F}_{12}$ or ${F}_{13}$) are assigned to the MDs associated with some SBS on both uplink and downlink access links.
\par
\noindent
\textit{\textbf{Assumption 9: }}Different subband slices of some subband (e.g., ${F}_{21}$, ${F}_{22}$ or ${F}_{23}$) are assigned to the MDs associated with some SBS on both uplink and downlink backhaul links.
\par
\noindent
\textit{\textbf{Assumption 10: }}Each file is transferred on some allocated subchannel (e.g., one subslice of some subband slice) at some cell (e.g., small cell or macrocell) on both downlink access and backhaul links.
\par
Significantly, Assumption 5 should be feasible since one file may be definitely divided into some blocks with same length. Assumption 6 shows the inter-cell interference between any two SBSs at each macrocell may be cancelled on both uplink and downlink access links; Assumption 7 similarly shows such a type of interference may also be cancelled on both uplink and downlink backhaul links. Assumption 8 shows the intra-cell interference between any two MDs at each picocell (SBS) may be cancelled on both uplink and downlink access links; Assumption 9 shows the intra-cell interference between any two MDs at each picocell may be cancelled on both uplink and downlink backhaul links; Assumption 10 shows the inter-file interference between any two MDs at each cell may be cancelled on both downlink access and backhaul links. In fact, the subfrequency slices in Assumptions 8-10 are the optimization parameters in this paper.
\par
Then, the downlink data rate $R_{nki}^{DL}$ of MD $k$ associated with SBS $n$ on subchannel $i$ can be given by
\begin{equation}\label{eq1}
R_{nki}^{DL}={{\beta }_{nki}}S_{n}^{DL}{{\log }_{2}}\Big( 1+\frac{p_{n}^{DL}{{h}_{nk}}}{{{\sigma }^{2}}} \Big)={{\beta }_{nki}}S_{n}^{DL}r_{nk}^{DL},
\end{equation}
where $\boldsymbol{\beta }=\left\{ {{\beta }_{nki}},\forall n\in {\Bar{\mathcal{N}}},\forall k\in {\Bar{\mathcal{K}}},\forall i\in \mathcal{I} \right\}$; ${{\beta }_{nki}}\in \left[ 0,1 \right]$ represents the frequency band fraction that is assigned to MD $k$ by SBS $n$ for transferring file $i$ on the downlink access link; $S_{n}^{DL}={a{W}{{T}_{2}}}/{3T{{M}}}$; ${{M}}$ is the number of SBSs at each macrocell; $p_{n}^{DL}$ represents the transmit power of SBS $n$; ${{h}_{nk}}$ denotes the channel gain between MD $k$ and SBS $n$; ${{\sigma }^{2}}$ is the noise power.
\par
Under the channel reciprocity, the uplink data rate $R_{nk}^{UL}$ of MD $k$ associated with SBS $n$ can be given by
\begin{equation}\label{eq2}
R_{nk}^{UL}={{\alpha }_{nk}}S_{n}^{UL}{{\log }_{2}}\Big( 1+\frac{p_{k}^{UL}{{h}_{nk}}}{{\sigma }^{2}} \Big)={{\alpha }_{nk}}S_{n}^{UL}r_{nk}^{UL},
\end{equation}
where $\boldsymbol{\alpha }=\big\{ {{\alpha }_{nk}},\forall n\in {\Bar{\mathcal{N}}},\forall k\in {\Hat{\mathcal{K}}} \big\}$; ${{\alpha }_{nk}}\in \left[ 0,1 \right]$ denotes the frequency band slice that is assigned to MD $k$ by SBS $n$; $S_{n}^{UL}={aW{{T}_{1}}}/{3T{{M}}}$; $p_{k}^{UL}$ represents the transmit power of MD $k$.
\par
Then, the downlink backhaul data rate of MD $k$ associated with SBS $n$ on subchannel $i$ can be given by
\begin{equation}\label{eq3}
R_{nki}^{BH}={{\eta }_{nki}}S_{n}^{BH}{{\log }_{2}}\Big( 1+\frac{p_{{{s}_{n}}}^{DL}{{h}_{{{s}_{n}}n}}}{{\sigma }^{2}} \Big)={{\eta }_{nki}}S_{n}^{BH}r_{n}^{BH},
\end{equation}
where $\boldsymbol{\eta }=\left\{ {{\eta }_{nki}},\forall n\in {\Bar{\mathcal{N}}},\forall k\in {\Bar{\mathcal{K}}},\forall i\in \mathcal{I} \right\}$; ${{\eta }_{nki}}\in \left[ 0,1 \right]$ represents the frequency band fraction that is assigned to MD $k$ by SBS $n$ for transferring file $i$ on the backhaul link; $S_{n}^{BH}={\left( 1-a \right)W{{T}_{2}}}/{3T{{M}}}$; $p_{{{s}_{n}}}^{DL}$ represents the transmit power of MBS ${{s}_{n}}$; ${{h}_{{{s}_{n}}n}}$ denotes the channel gain between SBS $n$ and MBS ${{s}_{n}}$. It is noteworthy that any SBS $n$ selects the nearest MBS ${{s}_{n}}$ to establish the corresponding backhaul link.
\subsection{Cache Model}
${{b}_{ni}}$ denotes the caching index of file $i$ at SBS $n$, it is equal to 1 if SBS $n$ has cached file $i$, 0 otherwise. In the reality, the file request often obeys some distribution (e.g., Zipf distribution), which can be attained by performing the statistical analysis of historical data stream. That is to say, the caching indices of files can be generated by obeying some distribution. When the required files have been cached at SBSs, HRDs can retrieve them from the SBSs selected by these MDs. Otherwise, HRDs can just attain the corresponding files from core network through backhaul links between SBSs and MBSs. In addition, ${{c}_{ki}}$ denotes the request index of HRD $k$ on the file $i$, it takes 1 if HRD $k$ requests file $i$, 0 otherwise.
\par
Then, the downlink access time (delay) that HRD $k$ spends on downloading the file $i$ from SBS $n$ can be given by
\begin{equation}\label{eq4}
t_{nki}^{DL}={{{{\Bar{d}}}_{i}}}/{R_{nki}^{DL}}={{{{\Bar{d}}}_{i}}}/{{{\beta }_{nki}}S_{n}^{DL}r_{nk}^{DL}},
\end{equation}
and the downlink backhaul time (delay) that HRD $k$ spends on downloading the file $i$ through the backhaul link between SBS $n$ and MBS ${{s}_{n}}$ can be given by
\begin{equation}\label{eq5}
t_{nki}^{BH}={{{{\Bar{d}}}_{i}}}/{R_{nki}^{BH}}={{{{\Bar{d}}}_{i}}}/{{{\eta }_{nki}}S_{n}^{BH}r_{n}^{BH}}.
\end{equation}
Thus, the time (delay), which is used for delivering file $i$ to HRD $k$ from SBS $n$, can be given by
\begin{equation}\label{eq6}
  t_{nki}^{HR}={{y}_{nk}}{{c}_{ki}}\left[ {{b}_{ni}}t_{nki}^{DL}+\left( 1-{{b}_{ni}} \right)\left( t_{nki}^{DL}+t_{nki}^{BH} \right) \right],
\end{equation}
where $t_{nki}^{DL}+t_{nki}^{BH}$ includes downlink access time and downlink backhaul time; ${{y}_{nk}}\in \left\{ 0,1 \right\}$ is the association index between HRD $k$ and SBS $n$; ${{y}_{nk}}=1$ if HRD $k$ is associated with SBS $n$, ${{y}_{nk}}=0$ otherwise; $\mathbf{Y}=\left\{ {{y}_{nk}},\forall n\in {\Bar{\mathcal{N}}},\forall k\in {\Bar{\mathcal{K}}} \right\}$.
\subsection{Computation Model}
As we know, any computation task of CSDs can be executed at mobile terminals or offloaded to SBSs. Assume that ${{d}_{k}}$ denotes the size of input data required by each computation task of CSD $k$, ${{C}_{k}}$ represents the number of CPU cycles required by the computation task of CSD $k$, $C_{n}^{ED}$ is the computation capability (i.e., the number of CPU cycles) of SBS $n$, and $C_{k}^{LC}$ denotes the one of CSD $k$. Then, the uplink access time (delay) that CSD $k$ spends on transmitting its computation task to SBS $n$ can be given by
\begin{equation}\label{eq7}
t_{nk}^{UL}={{{d}_{k}}}/{R_{nk}^{UL}}={{{d}_{k}}}/{{{\alpha }_{nk}}S_{n}^{UL}r_{nk}^{UL}},
\end{equation}
the execution time (delay) of computation task of CSD $k$ at SBS $n$ can be given by $t_{nk}^{ED}={{{C}_{k}}}/{{{\gamma }_{nk}}C_{n}^{ED}}$
and the local execution time (delay) of computation task of CSD $k$ can be given by $t_{k}^{LC}={{{C}_{k}}}/{C_{k}^{LC}}$,
where $\boldsymbol{\gamma }=\big\{ {{\gamma }_{nk}},\forall n\in {\Bar{\mathcal{N}}},\forall k\in {\Hat{\mathcal{K}}} \big\}$; ${{\gamma }_{nk}}\in \left[ 0,1 \right]$ represents the computing resource fraction occupied by CSD $k$ at SBS $n$.
\par
In general, the time (delay), which is used for completing the computation task of CSD $k$, can be given by
\begin{equation}\label{eq8}
t_{k}^{CS}=\sum\nolimits_{n\in \overline{\mathcal{N}}}{{{x}_{nk}}\left( t_{nk}^{UL}+t_{nk}^{ED} \right)}+t_{k}^{LC}\Big( 1-\sum\nolimits_{n\in \overline{\mathcal{N}}}{{{x}_{nk}}} \Big),
\end{equation}
where $t_{nk}^{UL}+t_{nk}^{ED}$ represents the time used for offloading the computation task from CSD $k$ to SBS $n$; ${{x}_{nk}}\in \left\{ 0,1 \right\}$ is the association index between CSD $k$ and SBS $n$; ${{x}_{nk}}=1$ if CSD $k$ is associated with SBS $n$, ${{x}_{nk}}=0$ otherwise; $\mathbf{X}=\big\{ {{x}_{nk}},\forall n\in {\Bar{\mathcal{N}}},\forall k\in {\Hat{\mathcal{K}}} \big\}$. As revealed in \eqref{eq8}, CSD $k$ completes the computation task locally when it is not associated with any SBS.
\section{Problem Formulation and Reformulation}
Next, we will provide some detailed descriptions on an optimization problem involved in this paper.
\subsection{Optimization Objective}
In small-cell IoT networks with mobile edge computing and caching, in order to minimize the sum of MDs' weighted delays under some limited resources, we try to perform joint MD association and resource allocation for HRDs and CSDs. Specifically, the objective function of our concentrated problem refers to the delay of HRDs and the one of CSDs. Mathematically, it can be given by
\begin{equation}\label{eq9}
  {\sum\nolimits_{k\in \Hat{\mathcal{K}}}{{{w}_{k}}t_{k}^{CS}}}+\sum\nolimits_{n\in \Bar{\mathcal{N}}}{\sum\nolimits_{k\in \Bar{\mathcal{K}}}{\sum\nolimits_{i\in \mathcal{I}}{{{w}_{k}}t_{nki}^{HR}}}},
\end{equation}
where the weighting parameter ${{w}_{k}}$ represents the priority of MD $k$.
\subsection{Necessary Constraints}
In this paper, we consider the single association for two types of MDs. In addition, some CSDs may not be associated with any SBS since they can execute their computation tasks locally. Thus, we have
\begin{equation}\label{eq10}
{{C}_{1}}:\sum\limits_{n\in \overline{\mathcal{N}}}{{{x}_{nk}}}\le 1,\forall k\in \widehat{\mathcal{K}},\  {{C}_{2}}:\sum\limits_{n\in \overline{\mathcal{N}}}{{{y}_{nk}}}=1,\forall k\in \overline{\mathcal{K}},
\end{equation}
where $\sum\nolimits_{n\in {\Bar{\mathcal{N}}}}{{{x}_{nk}}}=0$ means CSD $k$ completes its computation task locally.
\par
In a real system, the spectrum bandwidth is always limited. That is to say, the allocated bandwidth cannot be more than the total bandwidth on the uplink and downlink access links, and downlink backhaul link, and the corresponding constraints can be respectively given by
\begin{equation}\label{eq11}
\left\{ \begin{split}
  & {{C}_{3}}:\sum\nolimits_{k\in {\Hat{\mathcal{K}}}}{{{\alpha }_{nk}}{{x}_{nk}}}\le 1,\forall n\in {\Bar{\mathcal{N}}},\\
 & {{C}_{4}}:\sum\nolimits_{k\in {\Bar{\mathcal{K}}}}{\sum\nolimits_{i\in \mathcal{I}}{{{c}_{ki}}{{\beta }_{nki}}{{y}_{nk}}}}\le 1,\forall n\in {\Bar{\mathcal{N}}},\\
 & {{C}_{5}}:\sum\nolimits_{k\in {\Bar{\mathcal{K}}}}{\sum\nolimits_{i\in \mathcal{I}}{{{c}_{ki}}\left( 1-{{b}_{ni}} \right){{\eta }_{nki}}{{y}_{nk}}}}\le 1,\forall n\in {\Bar{\mathcal{N}}}. \\
\end{split} \right.
\end{equation}
In addition, any caching server has a limited storage capacity. That means the size of all contents cached at any SBS $n\in {\Bar{\mathcal{N}}}$ should not exceed such a capacity, i.e.,
\begin{equation}\label{eq12}
{{C}_{6}}:\sum\nolimits_{i\in \mathcal{I}}{{{b}_{ni}}{{{\Bar{d}}}_{i}}}+\sum\nolimits_{k\in {\Hat{\mathcal{K}}}}{{{d}_{k}}{{x}_{nk}}}\le {{D}_{n}}.
\end{equation}
Similarly, any computing server has a limited computing capability so that we cannot offload too many computing tasks to it simultaneously. That means the amount of all computing resources assigned to MDs by some computing server cannot exceed the computing capability of such a server, i.e.,
\begin{equation}\label{eq13}
{{C}_{7}}:\sum\nolimits_{k\in {\Hat{\mathcal{K}}}}{{{\gamma }_{nk}}{{x}_{nk}}}\le 1,\forall n\in {\Bar{\mathcal{N}}},
\end{equation}
where ${{\gamma }_{nk}}\in \left[ 0,1 \right]$ represents the computing resource fraction assigned to MD $k$ by the computing server of SBS $n$.
\par
To support any MD $k\in {\Bar{\mathcal{K}}}$ retrieving some required file $i\in \mathcal{I}$ from core network through backhaul links of some SBS $n\in {\Bar{\mathcal{N}}}$, the data rate on the downlink access link should be less than or equal to the one on the downlink backhaul link.
\begin{equation}\label{eq14}
{{C}_{8}}:{{c}_{ki}}\left( 1-{{b}_{ni}} \right){{y}_{nk}}R_{nki}^{DL}\le {{c}_{ki}}\left( 1-{{b}_{ni}} \right){{y}_{nk}}R_{nki}^{BH}.
\end{equation}
\subsection{Problem Formulation}
Under the resource constraints, an optimization problem with minimizing the sum of MDs' weighted delays, jointly performing the MD association and resource allocation in small-cell networks with mobile edge computing and caching, is formulated. Specifically, it can be given by
\begin{equation}\label{eq15}
\begin{split}
  \mathbf{P1:}&\underset{\mathbf{X},\mathbf{Y},\boldsymbol{\alpha },\boldsymbol{\beta },\boldsymbol{\gamma },\boldsymbol{\eta }}{\mathop{\min }}\,{F}\left( \mathbf{X},\mathbf{Y},\boldsymbol{\alpha },\boldsymbol{\beta },\boldsymbol{\gamma },\boldsymbol{\eta } \right) \\
 \text{s.t. }& {{C}_{1}},{{C}_{2}},{{C}_{3}},{{C}_{4}},{{C}_{5}},{{C}_{6}},{{C}_{7}},{{C}_{8}}, \\
 & {{C}_{9}}:{{x}_{nk}}\in \left\{ 0,1 \right\},\forall n\in {\Bar{\mathcal{N}}},\forall k\in {\Hat{\mathcal{K}}}, \\
 & {{C}_{10}}:{{y}_{nk}}\in \left\{ 0,1 \right\},\forall n\in {\Bar{\mathcal{N}}},\forall k\in {\Bar{\mathcal{K}}}, \\
 & {{C}_{11}}:0\le {{\alpha }_{nk}}\le 1,\forall n\in {\Bar{\mathcal{N}}},\forall k\in {\Hat{\mathcal{K}}}, \\
 & {{C}_{12}}:0\le {{\gamma }_{nk}}\le 1,\forall n\in {\Bar{\mathcal{N}}},\forall k\in {\Hat{\mathcal{K}}}, \\
 & {{C}_{13}}:0\le {{\beta }_{nki}}\le 1,\forall n\in {\Bar{\mathcal{N}}},\forall k\in {\Bar{\mathcal{K}}},\forall i\in \mathcal{I}, \\
 & {{C}_{14}}:0\le {{\eta }_{nki}}\le 1,\forall n\in {\Bar{\mathcal{N}}},\forall k\in {\Bar{\mathcal{K}}},\forall i\in \mathcal{I}, \\
\end{split}
\end{equation}
where
\begin{equation}\label{eq16}
\begin{split}
  & {F}\left( \mathbf{X},\mathbf{Y},\boldsymbol{\alpha },\boldsymbol{\beta },\boldsymbol{\gamma },\boldsymbol{\eta } \right) \\
  &=\sum\limits_{n\in \overline{\mathcal{N}}}{\sum\limits_{k\in \overline{\mathcal{K}}}{\sum\limits_{i\in \mathcal{I}}{{{w}_{k}}{{y}_{nk}}{{c}_{ki}}\left[ \left( 1-{{b}_{ni}} \right)\left( t_{nki}^{DL}+t_{nki}^{BH} \right)+{{b}_{ni}}t_{nki}^{DL} \right] }}} \\
  &\ \ \ \ +\sum\limits_{n\in {\Bar{\mathcal{N}}}}{\sum\limits_{k\in {\Hat{\mathcal{K}}}}{{{w}_{k}}{{x}_{nk}}\left( t_{nk}^{UL}+t_{nk}^{ED}-{t_{k}^{LC}} \right)}}+\sum\limits_{k\in {\Hat{\mathcal{K}}}}{{{w}_{k}}{t_{k}^{LC}}}.\\
\end{split}
\end{equation}
\subsection{Problem Reformulation}
It is easy to find that the MD association indicators $\left\{\mathbf{X},\mathbf{Y}\right\}$ and resource allocation parameters $\left\{\boldsymbol{\alpha },\boldsymbol{\beta },\boldsymbol{\gamma },\boldsymbol{\eta }\right\}$ are coupling in the problem \textbf{P1}. To decouple them, the alternating optimization may be well worth advocating. Consequently, the problem \textbf{P1} can be solved by alternatively optimizing the following two subproblems, i.e., problems \textbf{P2} and \textbf{P3}.
\begin{equation}\label{eq17}
\begin{matrix}
   \ \ \ \begin{split}
  & \textbf{P2: }\underset{\mathbf{X},\mathbf{Y}}{\mathop{\min }}\,F(\mathbf{X},\mathbf{Y}) \\
 & \text{s.t. }{{C}_{1}}\text{--}{{C}_{10}}, \\
\end{split} &\ \ \begin{split}
  & \textbf{P3: }\underset{\boldsymbol{\alpha },\boldsymbol{\beta },\boldsymbol{\gamma },\boldsymbol{\eta }}{\mathop{\min }}\,F(\boldsymbol{\alpha },\boldsymbol{\beta },\boldsymbol{\gamma },\boldsymbol{\eta }) \\
 & \text{s.t. }{{C}_{3}}\text{--}{{C}_{5}},{{C}_{7}}\text{--}{{C}_{8}},{{C}_{11}}\text{--}{{C}_{14}}. \\
\end{split}  \\
\end{matrix}
\end{equation}
\par
Evidently, the problem \textbf{P2} is in a decoupling form with respect to $\mathbf{X}$ and $\mathbf{Y}$. That is to say, such a problem can be further cut into
\begin{equation}\label{eq18}
\begin{matrix}
  \ \ \ \ \ \begin{split}
  & \textbf{P4: }\underset{\mathbf{X}}{\mathop{\min }}\,{F}\left( \mathbf{X} \right) \\
 & \text{s.t. }{{C}_{1}},{{C}_{3}},{{C}_{6}},{{C}_{7}}, \\
\end{split} &\ \ \ \ \ \ \ \begin{split}
  & \textbf{P5: }\underset{\mathbf{Y}}{\mathop{\min }}\,{F}\left( \mathbf{Y} \right) \\
 & \text{s.t. }{{C}_{2}},{{C}_{4}},{{C}_{5}},{{C}_{8}}. \\
\end{split}  \\
\end{matrix}
\end{equation}
\section{Algorithm Design}
To solve the problem \textbf{P1}, the subproblems \textbf{P3}, \textbf{P4} and \textbf{P5} are tackled alternately. Specifically, the general procedure, used for solving the problem \textbf{P1}, is illustrated in Algorithm 1. Such an algorithm is simply named as the MD association with minimizing network delay (AMND).
\begin{table}[h]
\centering
\begin{tabular}{ll}
\toprule[1pt]
\textbf{Algorithm 1:} AMND \\ \midrule[0.5pt]
1: \textbf{Initialization:} Run the initialization algorithms; ${t_1}=0$.\\
2: \textbf{Repeat (First Loop):}\\
3:\ \ \ \ Run the association subalgorithm for CSDs to solve problem \textbf{P4}.\\
4:\ \ \ \ Run the association subalgorithm for HRDs to solve problem \textbf{P5}.\\
5:\ \ \ \ Run the resource allocation subalgorithm to solve problem \textbf{P3}.\\
6:\ \ \ \ Update the iteration index: ${t_1}={t_1}+1$.\\
7: \textbf{Until} ${t_1}$ reaches ${T}_{1}$ iterations.\\ \bottomrule[0.5pt]
\end{tabular}
\label{tab1}
\end{table}
\par
Next, the optimization problems \textbf{P4} and \textbf{P5} are tackled at first. To this end, we first recall some definitions and results of coalitional game theorem \cite{LPQian2017Sep,JZhou2019}.
\subsection{Coalitional Game}
In a coalitional game, any MD has incentives to establish or join some coalition to reduce the sum of MDs' weighted delays, where some MDs associated with the same SBS form a coalition. Significantly, a virtual SBS is used for establishing a coalition in the CSD association, and all MDs who utilize the local computing resources join this coalition.
\par
To proceed, some necessary definitions \cite{JZhou2019,NZhao2019Oct} are given as follows.
\par
\noindent
\textit{\textbf{Definition 1: }}The (overall) transferable utility of HRDs is defined as the sum of weighted data delivery delay of HRDs, i.e.,
\begin{equation}\label{eq19}
 \Bar{\mathcal{V}}\left( \Bar{\mathfrak{U}} \right)=\sum\nolimits_{n\in \Bar{\mathcal{N}}}{\sum\nolimits_{k\in \Bar{\mathcal{K}}}{\sum\nolimits_{i\in \mathcal{I}}{{{w}_{k}}t_{nki}^{HR}}}},
\end{equation}
where $\Bar{\mathcal{V}}\left( \Bar{\mathfrak{U}} \right)$ still represents the payoff function with respect to partition $\Bar{\mathfrak{U}}=\left\{ {{\Bar{\mathcal{U}}}_{1}},{{\Bar{\mathcal{U}}}_{2}},\cdots ,{{\Bar{\mathcal{U}}}_{{\Bar{N}}}} \right\}$; ${{\Bar{\mathcal{U}}}_{n}}$ denotes the coalition (set) consisting of HRDs associated with SBS $n$.
\par
It is noteworthy that any two coalitions have empty intersection, and the union of all coalitions is the set of HRDs. That means
\begin{equation}\label{eq20}
\left\{ \begin{split}
  & {{\Bar{\mathcal{U}}}_{1}}\cup {{\Bar{\mathcal{U}}}_{2}}\cup \cdots \cup {{\Bar{\mathcal{U}}}_{\Bar{N}}}={\Bar{\mathcal{K}}}, \\
 & {{\Bar{\mathcal{U}}}_{m}}\cap {{\Bar{\mathcal{U}}}_{n}}=\phi,m\ne n, \\
\end{split} \right.
\end{equation}
\par
\noindent
\textit{\textbf{Definition 2: }}The (overall) transferable utility of CSDs is defined as the sum of weighted task computation delay of CSDs, i.e.,
\begin{equation}\label{eq21}
 \Hat{\mathcal{V}}\big( \Hat{\mathfrak{U}} \big)=\sum\nolimits_{n\in \Bar{\mathcal{N}}}{\sum\nolimits_{k\in \Hat{\mathcal{K}}}{{{w}_{k}}t_{nk}^{CS}}},
\end{equation}
where $\Hat{\mathcal{V}}\big( \Hat{\mathfrak{U}} \big)$ also denotes the payoff function with respect to partition $\Hat{\mathfrak{U}}=\big\{ {{\Hat{\mathcal{U}}}_{1}},{{\Hat{\mathcal{U}}}_{2}},\cdots ,{{\Hat{\mathcal{U}}}_{{\Bar{N}}}},{{\Hat{\mathcal{U}}}_{{\Bar{N}+1}}} \big\}$; ${{\Hat{\mathcal{U}}}_{n}}$ is the coalition (set) consisting of CSDs associated with SBS $n$, and ${{\Hat{\mathcal{U}}}_{{\Bar{N}+1}}}$ represents the one of CSDs associated with virtual SBS ${{\Bar{N}+1}}$.
\par
Significantly, any two coalitions have empty intersection, and the union of all coalitions is the set of CSDs. That means
\begin{equation}\label{eq22}
\left\{ \begin{split}
  & {{\Hat{\mathcal{U}}}_{1}}\cup {{\Hat{\mathcal{U}}}_{2}}\cup \cdots \cup {{\Hat{\mathcal{U}}}_{\Bar{N}}} \cup {{\Hat{\mathcal{U}}}_{\Bar{N}+1}}={\Hat{\mathcal{K}}}, \\
 & {{\Hat{\mathcal{U}}}_{m}}\cap {{\Hat{\mathcal{U}}}_{n}}=\phi,m\ne n, \\
\end{split} \right.
\end{equation}
\par
\noindent
\textit{\textbf{Definition 3: }}The coalitional game used for solving the problem \textbf{P4} or \textbf{P5} can be defined as $\left( {\Bar{\mathcal{K}}},\Bar{\mathfrak{U}},\Bar{\mathcal{V}} \right)$ or $\big( {\Hat{\mathcal{K}}},\Hat{\mathfrak{U}},\Hat{\mathcal{V}} \big)$  respectively.
\par
\noindent
\textit{\textbf{Definition 4: }}In the minimization optimization problem, the preference relation for any two partitions $\mathfrak{U}$ and $\tilde{\mathfrak{U}}$ can be defined as $\succ $. Specifically, $\mathfrak{U}\succ \tilde{\mathfrak{U}}$ denotes that $\tilde{\mathfrak{U}}$ is more preferable than $\mathfrak{U}$ for all players. It is noteworthy that they can represent the MD partitions of HRDs, or refer to the ones of CSDs.
\par
\noindent
\ding{172} If MD $i$ is in the coalition ${{\mathcal{U}}_{m}}$, then ${{\mathcal{U}}_{m}}{{\succ }^{i}}{{\mathcal{U}}_{n}}$ indicates that MD $i$ is willing to join the coalition ${{\mathcal{U}}_{n}}$.
\par
\noindent
\ding{173} If MDs $i$ and $j$ are in the coalitions ${{\mathcal{U}}_{m}}$ and ${{\mathcal{U}}_{n}}$ respectively, then ${{\mathcal{U}}_{m}}\succ _{j}^{i}{{\mathcal{U}}_{n}}$ indicates that MDs $i$ and $j$ are willing to exchange the coalitions.
\par
Mathematically, the relations stated in Definition 4 can be listed as follows.
\begin{equation}\label{eq23}
\mathfrak{U}\succ \tilde{\mathfrak{U}} \Leftrightarrow \mathcal{V}\left( \mathfrak{U} \right)>\mathcal{V}\big( {\tilde{\mathfrak{U}}} \big) ,
\end{equation}
\begin{equation}\label{eq24}
\left\{ \begin{split}
  & {{\mathcal{U}}_{m}}{{\succ }^{i}}{{\mathcal{U}}_{n}}\Leftrightarrow \mathcal{V}\left( {{\mathcal{U}}_{m}} \right)+\mathcal{V}\left( {{\mathcal{U}}_{n}} \right)>\mathcal{V}\big( {{{\tilde{\mathcal{U}}}}_{m}} \big)+\mathcal{V}\big( {{{\tilde{\mathcal{U}}}}_{n}} \big), \\
 & {{{\tilde{\mathcal{U}}}}_{m}}=\left\{ {{\mathcal{U}}_{m}}\backslash \left\{ i \right\} \right\}, \ \ \ {{{\tilde{\mathcal{U}}}}_{n}}=\left\{ \mathcal{U}{{}_{n}}\cup \left\{ i \right\} \right\}, \\
\end{split} \right.
\end{equation}
\begin{equation}\label{eq25}
\left\{ \begin{split}
  & {{\mathcal{U}}_{m}}\succ _{j}^{i}{{\mathcal{U}}_{n}} \Leftrightarrow \mathcal{V}\left( {{\mathcal{U}}_{m}} \right)+\mathcal{V}\left( {{\mathcal{U}}_{n}} \right)>\mathcal{V}\big( {{{\tilde{\mathcal{U}}}}_{m}} \big)+\mathcal{V}\big( {{{\tilde{\mathcal{U}}}}_{n}} \big), \\
 & {{{\tilde{\mathcal{U}}}}_{m}}=\left\{ {{\mathcal{U}}_{m}}\backslash \left\{ i \right\}\cup \left\{ j \right\} \right\}, \ \ \ {{{\tilde{\mathcal{U}}}}_{n}}=\left\{ \mathcal{U}{{}_{n}}\backslash \left\{ j \right\}\cup \left\{ i \right\} \right\}, \\
\end{split} \right.
\end{equation}
where $\tilde{\mathfrak{U}}=\big\{ \mathfrak{U}\backslash \left\{ {{\mathcal{U}}_{m}},{{\mathcal{U}}_{n}} \right\},{{{\tilde{\mathcal{U}}}}_{m}},{{{\tilde{\mathcal{U}}}}_{n}} \big\}$.
\par
\noindent
\textit{\textbf{Definition 5: }}A coalitional partition $\mathfrak{U}$ is Nash-stable if no further operations can be utilized to increase the transferable utility for any MD $i$.
\subsection{Solving the MD Association Subproblems}
According to the coalitional game theorem, we can easily develop a feasible algorithm to solve the problems \textbf{P4} or \textbf{P5}, which works as follows.
\begin{table}[h]
\centering
\begin{tabular}{ll}
\toprule[1pt]
\textbf{Algorithm 2:} ABCG \\ \midrule[0.5pt]
\textbf{ // Algorithm on the Sides of HRDs.}\\
1: Let any HRD select some SBS with strongest channel gain, \\
  \ \ \ which satisfies $ar_{nk}^{DL}\le \left( 1-a \right)r_{n}^{BH}$ simultaneously.\\
2: All MDs associated with some SBS $n$ form an initial coalition $\mathcal{U}_{n}^{0}$.\\
3: Employ the equal bandwidth allocation for associated HRDs on the\\
  \ \ \ downlink access and backhaul links, and establish  ${{\boldsymbol{\beta }}^{0}}$ and ${{\boldsymbol{\eta }}^{0}}$.\\
4: If MDs are not associated with any SBS, we let ${{\boldsymbol{\beta }}^{0}}$ and ${{\boldsymbol{\eta }}^{0}}$ take some\\
  \ \ \ small enough value (e.g., $1e^{-8}$).\\
5: Calculate the delays ${{\mathbf{t}}^{DL}}$ and ${{\mathbf{t}}^{BH}}$.\\
\textbf{ // Algorithm on the Sides of CSDs.}\\
6: Let any CSD select some SBS with strongest channel gain.\\
7: All MDs associated with some SBS $n$ form an initial coalition $\mathcal{U}_{n}^{0}$.\\
8: Employ the equal bandwidth allocation for associated CSDs on the \\
  \ \ \ uplink access links, and establish ${{\boldsymbol{\alpha }}^{0}}$.\\
9: Under the equal resource allocation, ${{\boldsymbol{\gamma }}^{0}}$ is established. \\
10: If MDs are not associated with any SBS, ${{\boldsymbol{\alpha }}^{0}}$ and ${{\boldsymbol{\gamma }}^{0}}$ take some small\\
  \ \ \ enough value (e.g., $1e^{-8}$).\\
11: Calculate the delays ${{\mathbf{t}}^{LC}}$, ${{\mathbf{t}}^{UL}}$ and ${{\mathbf{t}}^{ED}}$.\\
12: Any CSD $k$ associated with SBS $n$ executes its computation tasks\\
  \ \ \ locally when $t_{nk}^{UL}+t_{nk}^{ED}>t_{k}^{LC}$ and the constraint ${{C}_{6}}$ is satisfied for\\
  \ \ \  SBS $n$. It means $x_{nk}^{0}=0$ and $\mathcal{U}_{n}^{0}=\mathcal{U}_{n}^{0}\backslash\left\{ k \right\}$ for any $n$, $x_{{\Bar{N}+1}\ k}^{0}=1$\\
  \ \ \  and $\mathcal{U}_{\Bar{N}+1}^{0}=\mathcal{U}_{\Bar{N}+1}^{0} \cup\left\{ k \right\}$ for virtual SBS.\\ \bottomrule[0.5pt]
\end{tabular}
\label{tab2}
\end{table}
\par
\textbf{\ding{182} Initialization: }To solve the MD association subproblems \textbf{P4} and \textbf{P5} using coalitional game, the corresponding feasible initial partition needs to be found. To this end, the MD association with best channel gain and equal resource allocation is proposed, which is written as ABCG for short. The detailed procedure of ABCG can be found in Algorithm 2, and it is used for the initiation of Algorithm 1. Significantly, the achieved optimal MD partition in a coalitional game represents the optimal solution of one problem.
\par
In Algorithm 2, the Step 2 shows that the constraint ${{C}_{8}}$ is satisfied if $ar_{nk}^{DL}\le \left( 1-a \right)r_{n}^{BH}$ under the equal bandwidth allocation; the Steps 3, 8 and 9 guarantee that the constraints ${{C}_{3}}\text{--}{{C}_{5}}$ and ${{C}_{7}}$ are satisfied under the equal resource allocation; the Steps 4 and 10 reveal that some resources should not be allocated to MDs if they are not associated with any SBS, but the negligible resources are actually assigned to these MDs for the need of algorithm design (e.g., avoiding ``0/0"); the Step 12 lets CSDs decide the computation mode (i.e., edge or local computation) to minimize the network-wide weighted delay under the constraint ${{C}_{6}}$, and updates the corresponding association indicators and MD partitions.
\par
It is evident that joint MD association and resource allocation is completed under strict resource constraints. In addition, such an association algorithm is the further evolution of best channel association, which has been widely used for conventional cellular networks.
\par
\textbf{\ding{183} Update of Partition: }As revealed in Definition 3, the problem \textbf{P4} or \textbf{P5} has been formulated as coalitional game $\left( {\Bar{\mathcal{K}}},\Bar{\mathfrak{U}},\Bar{\mathcal{V}} \right)$ or $\big( {\Hat{\mathcal{K}}},\Hat{\mathfrak{U}},\Hat{\mathcal{V}} \big)$ respectively. According to the rule of such a game, a detailed framework will be designed to achieve the optimal solution of problem \textbf{P4} or \textbf{P5}, which is illustrated Algorithm 3. In the Step 3, two coalitions (e.g., ${{\mathcal{U}}_{m}}$ and ${{\mathcal{U}}_{n}}$) are randomly selected from the current partition. In the Steps 4-9, some randomly selected MD joins another empty coalition to establish two temporal coalitions (e.g., ${{\tilde{\mathcal{U}}}_{m}}$ and ${{\tilde{\mathcal{U}}}_{n}}$). In the Steps 11-13, the same kind of two randomly selected MDs are willing to exchange the coalitions and then establish two temporal coalitions. In the Steps 15-19, two temporal coalitions are made as current coalitions if they are feasible and satisfy $\mathcal{V}\big( {{{\tilde{\mathcal{U}}}}_{m}} \big)+\mathcal{V}\big( {{{\tilde{\mathcal{U}}}}_{n}} \big)<\mathcal{V}\left( {{\mathcal{U}}_{m}} \right)+\mathcal{V}\left( {{\mathcal{U}}_{n}} \right)$.
\begin{table}[h]
\centering
\begin{tabular}{ll}
\toprule[1pt]
\textbf{Algorithm 3:} MD Association Subalgorithm\\ \midrule[0.5pt]
1: \textbf{Initialization:} ${t_2}=0$.\\
2: \textbf{Repeat (Second Loop):}\\
3:\ \ \ \ ${{\mathcal{U}}_{m}}$ and ${{\mathcal{U}}_{n}}$ are randomly selected from the MD partition ${{\mathfrak{U}}^{{{t}_{2}}}}$.\\
4:\ \ \ \ \textbf{If} ${{\mathcal{U}}_{m}}$ is empty, \textbf{then}\\
5:\ \ \ \ \ \ MD $j$ is randomly selected from  ${{\mathcal{U}}_{n}}$. \\
6:\ \ \ \ \ \ Let ${{\tilde{\mathcal{U}}}_{m}}={{\mathcal{U}}_{m}}\cup \left\{ j \right\}$ and ${{\tilde{\mathcal{U}}}_{n}}={{\mathcal{U}}_{n}}\backslash \left\{ j \right\}$.\\
7:\ \ \ \ \textbf{ElseIf} ${{\mathcal{U}}_{n}}$ is empty, \textbf{then}\\
8:\ \ \ \ \ \ MD $i$ is randomly selected from ${{\mathcal{U}}_{m}}$. \\
9:\ \ \ \ \ \ Let ${{\tilde{\mathcal{U}}}_{m}}={{\mathcal{U}}_{m}}\backslash \left\{ i \right\}$ and ${{\tilde{\mathcal{U}}}_{n}}={{\mathcal{U}}_{n}}\cup \left\{ i \right\}$.\\
10:\ \ \ \textbf{Else} \\
11:\ \ \ \ \ The same type of MDs $i$  and $j$ are randomly selected from ${{\mathcal{U}}_{m}}$ and\\
12:\ \ \ \ \ ${{\mathcal{U}}_{n}}$ respectively. \\
13:\ \ \ \ \ Let ${{\tilde{\mathcal{U}}}_{m}}={{\mathcal{U}}_{m}}\backslash \left\{ i \right\}\cup \left\{ j \right\}$ and ${{\tilde{\mathcal{U}}}_{n}}={{\mathcal{U}}_{n}}\backslash \left\{ j \right\}\cup \left\{ i \right\}$.\\
14:\ \ \ \textbf{EndIf} \\
15:\ \ \ \textbf{If} ${{\tilde{\mathcal{U}}}_{m}}$ and ${{\tilde{\mathcal{U}}}_{n}}$ are feasible, \textbf{then}\\
16:\ \ \ \ \ \ \textbf{If} $\mathcal{V}\big( {{{\tilde{\mathcal{U}}}}_{m}} \big)+\mathcal{V}\big( {{{\tilde{\mathcal{U}}}}_{n}} \big)<\mathcal{V}\left( {{\mathcal{U}}_{m}} \right)+\mathcal{V}\left( {{\mathcal{U}}_{n}} \right)$, \textbf{then}\\
17:\ \ \ \ \ \ \ \ \ ${{\mathcal{U}}_{m}}={{\tilde{\mathcal{U}}}_{m}}$, ${{\mathcal{U}}_{n}}={{\tilde{\mathcal{U}}}_{n}}$.\\
18:\ \ \ \ \ \ \textbf{EndIf}\\
19:\ \ \ \textbf{EndIf}\\
20: \ \ Update the MD partition for HRDs: ${{\mathfrak{U}}^{{{t}_{2}}}}=\left\{ {{\mathcal{U}}_{1}},{{\mathcal{U}}_{2}},\cdots ,{{\mathcal{U}}_{\Bar{N}}} \right\}$. \\
21: \ \ Update the MD partition for CSDs: \\
22: \ \ \ \ \ \ \ \ \ \ \ \ \ \ \ \ \ \ ${{\mathfrak{U}}^{{{t}_{2}}}}=\left\{ {{\mathcal{U}}_{1}},{{\mathcal{U}}_{2}},\cdots ,{{\mathcal{U}}_{\Bar{N}}} ,{{\mathcal{U}}_{\Bar{N}+1}} \right\}$.\\
23:\ \ \ Update the iteration index: ${t_2}={t_2}+1$.\\
24: \textbf{Until} ${t_2}$ reaches ${T}_{2}$ iterations.\\ \bottomrule[0.5pt]
\end{tabular}
\label{tab3}
\end{table}
\par
Evidently, in order to select some proper SBSs, the HRDs can perform Algorithm 3 but skip Steps 21-22, the CSDs can execute Algorithm 3 but skip Step 20. Therefore, HRDs and CSDs can carry out their respective association subalgorithms in a parallel manner. Certainly, these MDs can also run their association subalgorithms in a centralized controller.
\subsection{Solving the Resource Allocation Subproblem}
Since the objective function of problem \textbf{P3} is convex and its constraints are linear, the problem \textbf{P3} should be a convex optimization problem. When ${{c}_{ki}}\left( 1-{{b}_{ni}} \right){{y}_{nk}}=1$, we have ${{\eta }_{nki}}\ge {a{{\beta }_{nki}}r_{nk}^{DL}}/{\left( 1-a \right)r_{n}^{BH}}$. To achieve some closed-form solutions of problem \textbf{P3}, we let ${{\eta }_{nki}}\ge \vartheta_{nk}= {ar_{nk}^{DL}}/{\left( 1-a \right)r_{n}^{BH}}$, which definitely meets the constraint $C_8$.
\par
According to the following results, we can easily solve the problem \textbf{P3}.
\par
\noindent
\textit{\textbf{Theorem 1: }}Under ${{c}_{ki}}{{y}_{nk}}=1$, ${{c}_{ki}}\left( 1-{{b}_{ni}} \right){{y}_{nk}}=1$ and ${{x}_{nk}}=1$, the closed-form solutions of $\boldsymbol{\alpha }$, $\boldsymbol{\beta }$, $\boldsymbol{\gamma }$ and $\boldsymbol{\eta }$ can be given by
\begin{equation}\label{eq26}
{{\alpha }_{nk}}={{\left[ \sqrt{Z_{nk}^{UL}/{{\Big( \sum\nolimits_{m\in \widehat{\mathcal{K}}}{{{x}_{nm}}\sqrt{Z_{nm}^{UL}}} \Big)}^{2}}\ } \right]}^{1}},
\end{equation}
\begin{equation}\label{eq27}
{{\beta }_{nki}}={{\left[ \sqrt{Z_{nki}^{DL}/{{\Big( \sum\nolimits_{m\in \overline{\mathcal{K}}}{\sum\nolimits_{i\in \mathcal{I}}{{{c}_{mi}}{{y}_{nm}}\sqrt{Z_{nmi}^{DL}}}} \Big)}^{2}}\ } \right]}^{1}},
\end{equation}
\begin{equation}\label{eq28}
{{\gamma }_{nk}}={{\left[ \sqrt{Z_{nk}^{ED}/{{\Big( \sum\nolimits_{m\in \widehat{\mathcal{K}}}{{{x}_{nm}}\sqrt{Z_{nm}^{ED}}} \Big)}^{2}}\ } \right]}^{1}},
\end{equation}
\begin{equation}\label{eq29}
{{\eta }_{nki}}=\left[ \sqrt{\frac{Z_{nki}^{BH}}{{\Big( \sum\limits_{m\in \overline{\mathcal{K}}}{\sum\limits_{i\in \mathcal{I}}{{{c}_{mi}}\left( 1-{{b}_{ni}} \right){{y}_{nm}}\sqrt{Z_{nmi}^{BH}}}} \Big)}^{2}}\ } \right]_{\vartheta }^{1},
\end{equation}
but they take a small enough value $\theta $ (e.g., $1e^{-8}$) under ${{x}_{nk}}=0$, ${{c}_{ki}}{{y}_{nk}}=0$ and ${{c}_{ki}}\left( 1-{{b}_{ni}} \right){{y}_{nk}}=0$, where $Z_{nk}^{UL}={{w}_{k}}{{{d}_{k}}}/{S_{n}^{UL}r_{nk}^{UL}}$; $Z_{nk}^{ED}={{w}_{k}}{{{C}_{k}}}/{C_{n}^{ED}}$; $Z_{nki}^{DL}={{w}_{k}}{{{{\Bar{d}}}_{i}}}/{S_{n}^{DL}r_{nk}^{DL}}$; $Z_{nki}^{BH}={{w}_{k}}{{{{\Bar{d}}}_{i}}}/{S_{n}^{BH}r_{n}^{BH}}$; $\left[ x \right]^{b}$ denotes $x=\min \left( x,b \right)$.
\par
\textit{Proof: }
We introduce the Lagrange multipliers $\boldsymbol{\lambda}=\left\{ {{{\lambda}}_{n}},\forall n\in {\Bar{\mathcal{N}}} \right\}$, $\boldsymbol{\mu }=\left\{ {{{\mu}}_{n}},\forall n\in {\Bar{\mathcal{N}}} \right\}$, $\boldsymbol{\nu}=\left\{ {{{\nu}}_{n}},\forall n\in {\Bar{\mathcal{N}}} \right\}$, and $\boldsymbol{\upsilon }=\left\{ {{{\upsilon }}_{n}},\forall n\in {\Bar{\mathcal{N}}} \right\}$ for the constraints ${{C}_{3}}\text{--}{{C}_{5}}$ and ${{C}_{7}}$ in the problem \textbf{P3} respectively. Then, the partial Lagrange function of problem \textbf{P3} with respect to these constraints can be given by
\begin{equation}\label{eq30}
\begin{split}
  & \mathcal{L}\left( \boldsymbol{\alpha },\boldsymbol{\beta },\boldsymbol{\gamma },\boldsymbol{\eta },\boldsymbol{\lambda },\boldsymbol{\mu },\boldsymbol{\upsilon },\boldsymbol{\nu } \right) \\
 & =\sum\limits_{n\in \bar{\mathcal{N}}}{\sum\limits_{k\in \bar{\mathcal{K}}}{\sum\limits_{i\in \mathcal{I}}{{{c}_{ki}}{{y}_{nk}}\left[ \left( 1-{{b}_{ni}} \right)\left( \frac{Z_{nki}^{DL}}{{{\beta }_{nki}}}+\frac{Z_{nki}^{BH}}{{{\eta }_{nki}}} \right) \right.}}} \\
 &\ \ \ \left. +\frac{{{b}_{ni}}Z_{nki}^{DL}}{{{\beta }_{nki}}} \right]+\sum\limits_{n\in \bar{\mathcal{N}}}{\sum\limits_{k\in \hat{\mathcal{K}}}{{{x}_{nk}}\left( \frac{Z_{nk}^{UL}}{{{\alpha }_{nk}}}+\frac{Z_{nk}^{ED}}{{{\gamma }_{nk}}} \right)}} \\
 &\ \ \  +\sum\limits_{n\in \bar{\mathcal{N}}}{{{\nu }_{n}}\Big( \sum\limits_{k\in \bar{\mathcal{K}}}{\sum\limits_{i\in \mathcal{I}}{{{c}_{ki}}\left( 1-{{b}_{ni}} \right){{\eta }_{nki}}{{y}_{nk}}}}-1 \Big)} \\
 &\ \ \  +\sum\limits_{n\in \bar{\mathcal{N}}}{{{\mu }_{n}}\Big( \sum\limits_{k\in \bar{\mathcal{K}}}{\sum\limits_{i\in \mathcal{I}}{{{c}_{ki}}{{\beta }_{nki}}{{y}_{nk}}}}-1 \Big)} \\
 &\ \ \  +\sum\limits_{n\in \bar{\mathcal{N}}}{{{\lambda }_{n}}\Big( \sum\limits_{k\in \hat{\mathcal{K}}}{{{\alpha }_{nk}}{{x}_{nk}}}-1 \Big)} \\
 &\ \ \  +\sum\limits_{n\in \bar{\mathcal{N}}}{{{\upsilon }_{n}}\Big( \sum\limits_{k\in \hat{\mathcal{K}}}{{{\gamma }_{nk}}{{x}_{nk}}}-1 \Big)}.\\
\end{split}
\end{equation}
\par
We let ${\partial \mathcal{L}}/{\partial {{\alpha }_{nk}}}=0$, ${\partial \mathcal{L}}/{\partial {{\lambda }_{n}}}=0$, ${\partial \mathcal{L}}/{\partial {{\gamma }_{nk}}}=0$ and ${\partial \mathcal{L}}/{\partial {{\upsilon }_{n}}}=0$ for any SBS $n$ and CSD $k$, ${\partial \mathcal{L}}/{\partial {{\beta }_{nki}}}=0$, ${\partial \mathcal{L}}/{\partial {{\mu }_{n}}}=0$, ${\partial \mathcal{L}}/{\partial {{\eta }_{nki}}}=0$ and ${\partial \mathcal{L}}/{\partial {{\nu }_{n}}}=0$ for any file $i$, SBS $n$ and HRD $k$, then have
\begin{equation}\label{eq31}
{{\alpha }_{nk}}=\left\{ \begin{split}
  & \sqrt{{Z_{nk}^{UL}}/{{{\lambda }_{n}}}},\text{  if  }{{x}_{nk}}=1, \\
 & \text{        }\theta,\text{       otherwise,} \\
\end{split} \right.
\end{equation}
\begin{equation}\label{eq32}
\sum\nolimits_{k\in {\hat{\mathcal{K}}}}{{{\alpha }_{nk}}{{x}_{nk}}}=1,
\end{equation}
\begin{equation}\label{eq33}
{{\beta }_{nki}}=\left\{ \begin{split}
  & \sqrt{{Z_{nki}^{DL}}/{{{\mu }_{n}}}},\text{  if  }{{c}_{ki}}{{y}_{nk}}=1, \\
 & \text{        }\theta,\text{       otherwise,} \\
\end{split} \right.
\end{equation}
\begin{equation}\label{eq34}
\sum\nolimits_{k\in \bar{\mathcal{K}}}{\sum\nolimits_{i\in \mathcal{I}}{{{c}_{ki}}{{\beta }_{nki}}{{y}_{nk}}}}=1,
\end{equation}
\begin{equation}\label{eq35}
{{\gamma }_{nk}}=\left\{ \begin{split}
  & \sqrt{{Z_{nk}^{ED}}/{{{\upsilon }_{n}}}},\text{  if  }{{x}_{nk}}=1, \\
 & \text{        }\theta,\text{       otherwise,} \\
\end{split} \right.
\end{equation}
\begin{equation}\label{eq36}
\sum\nolimits_{k\in \hat{\mathcal{K}}}{{{\gamma }_{nk}}{{x}_{nk}}}=1,
\end{equation}
\begin{equation}\label{eq37}
{{\eta }_{nki}}=\left\{ \begin{split}
  & \sqrt{{Z_{nki}^{BH}}/{{{\nu }_{n}}}},\text{  if  }{{c}_{ki}}\left( 1-{{b}_{ni}} \right){{y}_{nk}}=1, \\
 & \text{        }\theta,\text{       otherwise,} \\
\end{split} \right.
\end{equation}
\begin{equation}\label{eq38}
\sum\nolimits_{k\in \bar{\mathcal{K}}}{\sum\nolimits_{i\in \mathcal{I}}{{{c}_{ki}}\left( 1-{{b}_{ni}} \right){{\eta }_{nki}}{{y}_{nk}}}}=1,
\end{equation}
where $\theta $ takes a small enough value (e.g., $1e^{-8}$) to avoid ``0/0".
\par
By combining \eqref{eq31} and \eqref{eq32}, we can achieve ${{\alpha }_{nk}}=\theta$ if ${{x}_{nk}}=0$, and have \eqref{eq26} if ${{x}_{nk}}=1$; under \eqref{eq33} and \eqref{eq34}, we can attain ${{\beta }_{nki}}=\theta$ if ${{c}_{ki}}{{y}_{nk}}=0$, and have \eqref{eq27} if ${{c}_{ki}}{{y}_{nk}}=1$; under \eqref{eq35} and \eqref{eq36}, we can obtain ${{\gamma }_{nk}}=\theta$ if ${{x}_{nk}}=0$, and have \eqref{eq28} if ${{x}_{nk}}=1$; according to \eqref{eq37} and \eqref{eq38}, we can easily get ${{\eta }_{nki}}=\theta$ if ${{c}_{ki}}\left( 1-{{b}_{ni}} \right){{y}_{nk}}=0$, and have \eqref{eq29} if ${{c}_{ki}}\left( 1-{{b}_{ni}} \right){{y}_{nk}}=1$.  \ding{113}
\par
Now, the whole procedure used for solving the problem \textbf{P3} can be found in Algorithm 4.
\begin{table}[h]
\centering
\begin{tabular}{ll}
\toprule[1pt]
\textbf{Algorithm 4:} Resource Allocation Subalgorithm \\ \midrule[0.5pt]
\textbf{// Algorithm on Sides of CSDs}\\
1: Calculate ${{\boldsymbol{\alpha }}}$ using \eqref{eq26}.\\
2: Calculate ${{\boldsymbol{\gamma }}}$ using \eqref{eq27}.\\
\textbf{// Algorithm on Sides of HRDs}\\
3: Calculate ${{\boldsymbol{\beta }}}$ using \eqref{eq28}.\\
4: Calculate ${{\boldsymbol{\eta }}}$ using \eqref{eq29}.\\ \bottomrule[0.5pt]
\end{tabular}
\label{tab4}
\end{table}
\section{Algorithm Analysis}
In this section, we mainly concentrate on the computation complexity and convergence of designed algorithms, and the stability of coalitional game.
\subsection{Convergence Analysis}
In general, Algorithm 1 (i.e., First Loop) based on alternating optimization should converge when Algorithms 3 and 4 (i.e., Second and Third Loops) converge. Next, we will prove the convergence of such these algorithms.
\par
\noindent
\textit{\textbf{Theorem 2: }}Algorithm 3 is convergent after some iterations.
\par
\textit{Proof: }Through a direct observation, we can easily find that the objective function and constraints of problems \textbf{P4} and \textbf{P5} are separable with respect to SBSs (coalitions). That is to say, the splitting, merging or exchanging operation on any two coalitions has not any impact on the transferable utility (i.e., weighted delay) achieved by MDs in other coalitions. Therefore, the overall transferable utility can be reduced if such these operations on any two coalitions can decrease the transferable utility of operated MDs in them.
\par
Seen from Steps 16-17, it is easy to find that the splitting, merging or exchanging operation on any two coalitions can decrease the transferable utility of operated MDs in them at each iteration. Evidently, Algorithm 3 always tries to minimize the sum of MDs' weighted delays under resource constraints. Therefore, when no further operations on the MD partition need to be made for reducing overall transferable utility, Algorithm 3 is thought to be convergent.
 \ding{113}

 \subsection{Stability Analysis}
 In this section, the stability analysis of coalitional game will be made, which is similar to the efforts in \cite{JZhou2019,NZhao2019Oct}.
 \par
\noindent
\textit{\textbf{Theorem 3: }}The final partition $\mathfrak{U}$ found by Algorithm 3 is Nash-stable.
\par
\textit{Proof: }As shown in Algorithm 3, the MD partition should be either updated or stay the same after each iteration. Since the limited number of coalitions often means that the maximal number of coalition structures is limited, the update of MD partition will be terminated after some splitting, merging or exchanging operations. When a coalitional game finally gets into a stable state $\mathfrak{U}$, no MDs will try to leave its current coalition. To prove such a stability of coalitional game, the counter-evidence-based method is utilized. Assume that the final MD partition attained by Algorithm 3 is unstable. There may exist some MD $i$, its coalition ${{\mathcal{U}}_{m}}$ and another distinct coalition ${{\mathcal{U}}_{n}}$ such that ${{\mathcal{U}}_{m}}{{\succ }^{i}}{{\mathcal{U}}_{n}}$. Certainly, there may still exist two different MDs $i$ and $j$, and their corresponding coalitions ${{\mathcal{U}}_{m}}$ and ${{\mathcal{U}}_{n}}$ such that ${{\mathcal{U}}_{m}}\succ _{j}^{i}{{\mathcal{U}}_{n}}$. Seen from Definitions 1 and 2, we can easily know that the decrement or increment of transferable utility of any MD can result in the corresponding change of the (overall) transferable utility of all MDs. That means one of splitting, merging and exchanging operations should be made definitely. Evidently, it is contrary to the statement that $\mathfrak{U}$ is a final partition.
\par
In general, the final partition $\mathfrak{U}$ achieved by Algorithm 3 is a Nash-stable structure.
 \ding{113}
\subsection{Complexity Analysis}
In this section, the computation complexity analyses of designed algorithms will be made.
\par
\textbf{\textit{1) The Computation Complexity of Algorithm 2}}
\par
Through a direct observation, we can easily find that Algorithm 2 is not in an iterative form. In addition, since the subalgorithms of joint MD association and resource allocation, designed for HRDs and CSDs, are not in a coupling form, they can be executed separately. That is to say, such these two subalgorithms for HRDs and CSDs can be performed in a parallel manner.
\par
In Algorithm 2, the computation complexity of subalgorithm of joint MD association and resource allocation for HRDs is mainly dependent on the calculation of ${{\boldsymbol{\beta }}^{0}}$ and ${{\boldsymbol{\eta }}^{0}}$. Therefore, such a computation complexity can be given by $\mathcal{O}\big({\Bar{N}}{\Bar{K}}I \big)$. In the subalgorithm of joint MD association and resource allocation for CSDs, most steps have a computation complexity of $\mathcal{O}\big( {\Bar{N}}{\Hat{\mathcal{K}}} \big)$, and other steps that have  a higher computation complexity than them don't exist.
\par
In general, the computation complexity of Algorithm 2 may be $\mathcal{O}\big( \max \big( {\Bar{N}}{\Hat{\mathcal{K}}},{\Bar{N}}{\Bar{K}}I \big) \big)$.
\par
\textbf{\textit{2) The Computation Complexity of Algorithm 3}}
\par
It is easy to find that the association subalgorithms for HRDs and CSDs are not coupling, and thus they can be performed separately. That means such these two subalgorithms can be executed concurrently.
\par
At first, we concentrate on the calculation of Step 15 in Algorithm 3. In the best case, the computation complexity of Step 15 is $\mathcal{O}\big( 1 \big)$ for any kind of MDs if an infeasible constraint is found from the beginning of condition judgment. In the worst case, it should be $\mathcal{O}\big({\Bar{K}} I \big)$ or $\mathcal{O}\big(\max\big({\Hat{K}}, I \big)\big)$ for HR or CSDs respectively if an infeasible constraint is found at the end of condition judgment. As for other steps in Algorithm 3, we can easily find that the computation complexity of them is mainly dependent on the MD selection. In the best case, the computation complexity of selection operation is $\mathcal{O}\big( 1 \big)$ for any kind of MDs when only one MD exists in any operated coalition. In the worst case, it should be $\mathcal{O}\big({\Bar{K}} \big)$ for HRDs when ${\Bar{K}}$ MDs are in some operated coalition, and it should be $\mathcal{O}\big({\Hat{K}}\big)$ for CSDs when ${\Hat{K}}$ MDs are in some operated coalition.
\par
After $T_2$ iterations, the computation complexity of Algorithm 3 may be $\mathcal{O}({{T}_{2}})$ in the best case, and $\mathcal{O}(\max (\Bar{K}I{{T}_{2}},\Hat{K}{{T}_{2}}))$ in the worst case.
\par
\textbf{\textit{3) The Computation Complexity of Algorithm 4}}
\par
Evidently, the update of $\boldsymbol{\alpha }$ and $\boldsymbol{\gamma }$ just refers to CSDs; the one of $\boldsymbol{\beta }$ and $\boldsymbol{\eta }$ just refers to HRDs. That is to say, such these two blocks used for updating parameters can be tackled separately, which means they can be executed in a parallel manner.
\par
In Algorithm 3, the computation complexity of all steps of subalgorithm used for updating $\boldsymbol{\alpha }$ and $\boldsymbol{\gamma }$ is $\mathcal{O}\big({\Bar{N}}{\Hat{K}} \big)$ for CSDs; the one of subalgorithm used for updating $\boldsymbol{\beta }$ and $\boldsymbol{\eta }$ may be $\mathcal{O}\big( {\Bar{N}}{\Bar{K}}I \big)$ for HRDs.
\par
In general, the computation complexity of Algorithm 3 may be $\mathcal{O}\big( \max \big({\Bar{N}}{\Hat{K}},{\Bar{N}}{\Bar{K}}I \big) \big)$.
\par
\textbf{\textit{4) The Computation Complexity of Algorithm 1}}
\par
By combining Algorithms 2, 3 and 4, we can easily find that the computation complexities of subalgorithms on the HRDs' and CSDs' sides in Algorithm 1 may be given by $\mathcal{O}\big( \max \big( {{T}_{1}}{{T}_{2}}, {\Bar{N}} \Bar{K} I{{T}_{1}} \big) \big)$ and $\mathcal{O}\big( \max \big( {{T}_{1}}{{T}_{2}}, {\Bar{N}} \Hat{K}{{T}_{1}}  \big) \big)$ respectively in the best case, and they may be given by $\mathcal{O}\big( \max \big(\Bar{K} I{{T}_{1}}{{T}_{2}}, {\Bar{N}} \Bar{K} I{{T}_{1}}  \big) \big)$ and $\mathcal{O}\big( \max \big(I{{T}_{1}}{{T}_{2}},\Hat{K}{{T}_{1}}{{T}_{2}}, {\Bar{N}} \Hat{K}{{T}_{1}} \big) \big)$ respectively in the worst case.
\par
In general, the computation complexity of Algorithm 1 may be $\mathcal{O}\big( \max \big( {{T}_{1}}{{T}_{2}}, {\Bar{N}} \Bar{K} I{{T}_{1}}, {\Bar{N}} \Hat{K}{{T}_{1}}   \big) \big)$ in the best case, and $\mathcal{O}\big( \max \big( \Bar{K} I{{T}_{1}}{{T}_{2}},\Hat{K}{{T}_{1}}{{T}_{2}},{\Bar{N}} \Bar{K} I{{T}_{1}}, {\Bar{N}} \Hat{K}{{T}_{1}} \big) \big)$ in the worst case.
\newsavebox\pra
\begin{lrbox}{\pra}
  \begin{minipage}{0.3\textwidth}
  \vspace{-0.7em}
    \begin{align*}
      {{\mathcal{P}}_\text{LOS}}\left( d \right)=&\big( 1-{{e}^{{-d}/{63}}} \big)\min \big( {18}/{d},1 \big)\\
      &+{{e}^{{-d}/{63}}} \text{ \small{[dB]}}
    \end{align*}
  \end{minipage}
\end{lrbox}
\newsavebox\prb
\begin{lrbox}{\prb}
  \begin{minipage}{0.3\textwidth}
  \vspace{-0.7em}
    \begin{align*}
     {{\mathcal{P}}_\text{LOS}}\left( d \right)=&\big( 1-{{e}^{{-d}/{72}}} \big)\min \big( {18}/{d},1 \big)\\
     &+{{e}^{{-d}/{72}}} \text{ \small{[dB]}}
    \end{align*}
  \end{minipage}
\end{lrbox}
\newsavebox\prc
\begin{lrbox}{\prc}
  \begin{minipage}{0.3\textwidth}
  \vspace{-0.7em}
    \begin{align*}
      {{\mathcal{P}}_\text{LOS}}\left( d \right)=&0.5-\min \big( 0.5,5{{e}^{{-156}/{d}}} \big)\\
      &+\min \big( 0.5,5{{e}^{{-d}/{30}}} \big) \text{ \small{[dB]}}
    \end{align*}
  \end{minipage}
\end{lrbox}

\begin{table}[]
\centering
\caption{SIMULATION PARAMETERS}
\begin{tabular}{|c|c|c|c|}
\hline  \rule{0pt}{8pt}
\textbf{Parameter}  &\textbf{Value} &\textbf{Parameter}  &\textbf{Value}\\ [1pt]
\hline \rule{0pt}{8pt}
Inter-site distance  &1000 m  &${p_{n}^{DL}}$ for MBS $n$& 46 dBm \\ [1pt]
\hline \rule{0pt}{8pt}
System bandwidth  & 20 MHz &${p_{n}^{DL}}$ for SBS $n$& 24 dBm\\ [1pt]
\hline \rule{0pt}{8pt}
Number of files $I$  & 20 &${p_{k}^{UL}}$ for MD $k$ & 23 dBm\\ [1pt]
\hline \rule{0pt}{8pt}
${{d}_{k}}$  & 100 KB &$C_k$ & $1 {e^{9}}$ cycles\\ [1pt]
\hline \rule{0pt}{8pt}
$C_{n}^{ED}$  & $6 {{e}^{10}}$ cycles &${C_{k}^{LC}}$ & $1.4 {{e}^{9}}$ cycles\\ [1pt]
\hline \rule{0pt}{8pt}
$\Bar{d}_{i}$  & 5 MB &$D_n$ & 2 GB\\ [1pt]
\hline
 \multicolumn{1}{|c|}{\rule{0pt}{8pt}\textbf{Parameter}} & \multicolumn{3}{c|}{\rule{0pt}{8pt}\textbf{Value}} \\ [1pt]
\hline
 \multicolumn{1}{|c|}{\rule{0pt}{8pt} Noise PSD} & \multicolumn{3}{c|}{\rule{0pt}{8pt} -174 dBm/Hz } \\ [1pt]
\hline
\multirow{4}{*}{\rule{0pt}{36pt} MBS-MD pathloss} &
\multicolumn{3}{c|}{\rule{0pt}{8pt}${{\ell}_\text{LOS}}\left( d \right)=30.8+24.2{{\log }_{10}}\left( d \right)$ \text{ [dB]}}\\ [1pt]
\cline{2-4}
&\multicolumn{3}{c|}{\rule{0pt}{8pt}${{\ell}_\text{NLOS}}\left( d \right)=2.7+42.8{{\log }_{10}}\left( d \right)$  \text{ [dB]}}\\ [1pt]
\cline{2-4}
&\multicolumn{3}{c|}{\usebox{\pra}}\\ [1pt]
\cline{2-4}
&\multicolumn{3}{c|}{\rule{0pt}{8pt}${{\Theta }_\text{sd}}=6$ \text{ dB}}\\ [1pt]
\hline
\multirow{4}{*}{\rule{0pt}{36pt}MBS-SBS pathloss} &
\multicolumn{3}{c|}{\rule{0pt}{8pt}${{\ell }_\text{LOS}}\left( d \right)=30.2+23.5{{\log }_{10}}\left( d \right)$  \text{ [dB]}}\\ [1pt]
\cline{2-4}
&\multicolumn{3}{c|}{\rule{0pt}{8pt}${{\ell }_\text{NLOS}}\left( d \right)=16.3+36.3{{\log }_{10}}\left( d \right)$  \text{ [dB]}}\\ [1pt]
\cline{2-4}
&\multicolumn{3}{c|}{\usebox{\prb}}\\ [1pt]
\cline{2-4}
&\multicolumn{3}{c|}{\rule{0pt}{8pt}${{\Theta }_\text{sd}}=6$ \text{ dB}} \\ [1pt]
\hline
\multirow{4}{*}{\rule{0pt}{36pt}SBS-MD pathloss} &
\multicolumn{3}{c|}{\rule{0pt}{8pt}${{\ell }_\text{LOS}}\left( d \right)=41.1+20.9{{\log }_{10}}\left( d \right)$  \text{ [dB]}}\\ [1pt]
\cline{2-4}
&\multicolumn{3}{c|}{\rule{0pt}{8pt}${{\ell }_\text{NLOS}}\left( d \right)=32.9+37.5{{\log }_{10}}\left( d \right)$  \text{ [dB]}}\\ [1pt]
\cline{2-4}
&\multicolumn{3}{c|}{\usebox{\prc}}\\ [1pt]
\cline{2-4}
&\multicolumn{3}{c|}{\rule{0pt}{8pt}${{\Theta }_\text{sd}}\left( d \right)=6\text{ dB }\left( \text{LOS} \right)$;  ${{\Theta }_\text{sd}}=4\text{ dB }\left( \text{NLOS} \right)$}\\ [1pt]
\hline
\end{tabular}
\label{tab5}
\end{table}
\section{PERFORMANCE EVALUATION}
In HCNs with mobile edge computing and caching, MBSs are deployed according to a cellular frame, SBSs and two types of MDs are randomly deployed into each macrocell. Some essential parameters are listed in TABLE \ref{tab5}, where the inter-site distance denotes the one between two MBSs; $d$ represents the MBS-MD, MBS-SBS or SBS-MD distance; PSD is the abbreviation of power spectral density; ${{\ell }_\text{LOS}}$ and ${{\ell }_\text{NLOS}}$ denote the line-of-sight (LOS) and non-line-of-sight (NLOS) pathloss respectively \cite{JHoydis2013}; ${{\mathcal{P}}_\text{LOS}}$ denotes the LOS probability \cite{JHoydis2013}; ${{\Theta }_\text{sd}}$ represents the standard deviation of shadowing fading \cite{JHoydis2013}.
\par
We assume that the popularity probability of file $i\in \mathcal{I}$ follows Zipf distribution \cite{LZhang2016Jun}, and it can be given by
\begin{equation}\label{eq39}
{{\Pr }_{i}}=\frac{{1}/{{{i}^{\delta }}}}{\sum\nolimits_{j\in \mathcal{I}}{{1}/{{{j}^{\delta }}}}},
\end{equation}
where $i$ represents the file index; $\delta$ represents the file request coefficient, and controls the popularity distribution of files. In general, the $c_{ki}$ for any user $k$ and file $i$ can be known before offloading design in reality. In order to catch traffic distribution well in a practical system, we assume that the file request probability of MDs is decided by the popularity probability of files in the simulation. That is to say, $c_{ki}$ is randomly generated for any user $k$ and file $i$ under the file request probability. At last, $b_{ni}$ for any SBS $n$ and file $i$ is established under the file request index and caching capacity limitation.
\par
In the simulation, we can easily find that the total time (delay) of HRDs may decrease with file request coefficient $\delta $. That because a high $\delta $ often means a high caching ratio. That is to say, a high $\delta $ may mean MDs have more opportunities to retrieve files from SBSs directly. In addition, AMND almost always achieves a lower average time (delay) than ABCG. The reason for this may be that the former mechanism always tries to minimize the sum of MDs' weighted delays, but the latter one just focuses on channel gains.
\begin{figure}[!t]
\centering
\centerline{\includegraphics[width=3.5in,height=3in]{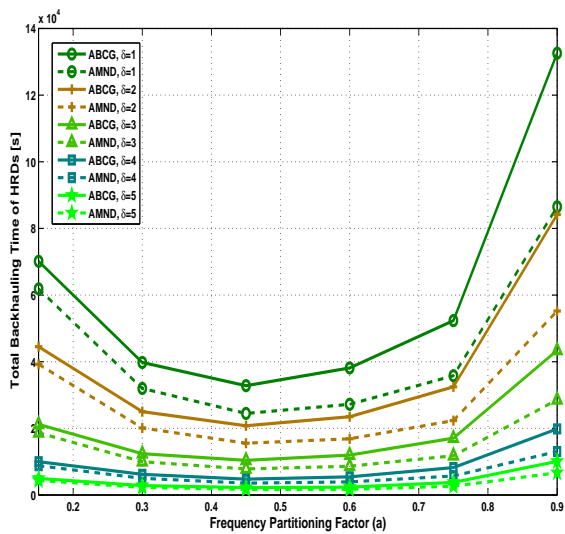}}
\caption{The total (downlink) backhauling time of HRDs versus frequency partitioning factor $a$ under the file request coefficient $\delta $.}
\label{fig4}
\end{figure}
\par
Fig. \ref{fig4} shows the impacts of frequency partitioning factor $a$ on the total (downlink) backhauling time (delay) of HRDs under the file request coefficient $\delta $. Similarly, in the low domain of $a$, an increased $a$ may mean that HRDs have more and more opportunities to get the services of some SBSs with high downlink data rates, and fewer and fewer HRDs may retrieve required files through backhauling links. However, in the high domain of $a$, under the constraint $C_8$, an increased $a$ may imply that HRDs have fewer and fewer opportunities to obtain the services of such these SBSs, and more and more HRDs may retrieve required files through backhauling links. Therefore, the total backhauling time of HRDs in ABCG and AMND may initially decrease but then increase with $a$.
\begin{figure}[!t]
\centering
\centerline{\includegraphics[width=3.5in,height=3in]{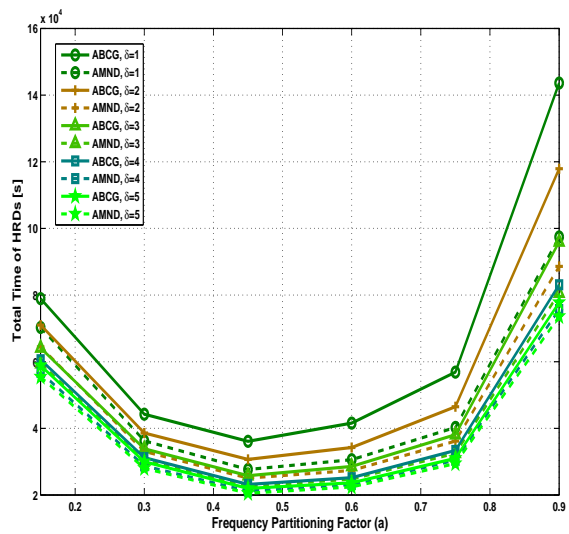}}
\caption{The total time of HRDs versus frequency partitioning factor $a$ under the file request coefficient $\delta $.}
\label{fig5}
\end{figure}
\par
Fig. \ref{fig5} shows the impacts of frequency partitioning factor $a$ on the total time (delay) of HRDs under the file request coefficient $\delta $. Evidently, in the low domain of $a$, an increased $a$ may mean that HRDs have more and more opportunities to access some SBSs with high downlink data rates. However, in the high domain of $a$, under the constraint $C_8$, an increased $a$ may imply that HRDs have fewer and fewer opportunities to access such these SBSs. Therefore, the total time of HRDs in ABCG and AMND may initially decrease but then increase with $a$.
\begin{figure}[!t]
\centering
\centerline{\includegraphics[width=3.5in,height=3in]{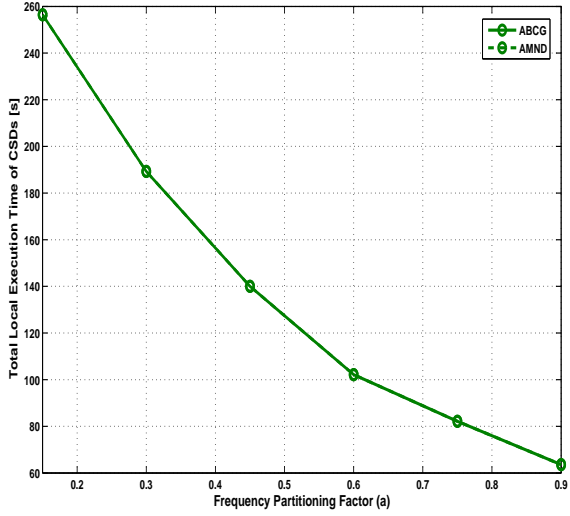}}
\caption{The total local computing time of CSDs versus frequency partitioning factor $a$ under the file request coefficient $\delta $.}
\label{fig6}
\end{figure}
\par
Fig. \ref{fig6} shows the impacts of frequency partitioning factor $a$ on the total local computing (execution) time (delay) of CSDs under the file request coefficient $\delta $. Evidently, an increased $a$ often means that CSDs have higher uplink data rates, and more and more CSDs may offload their computing tasks to SBSs. Therefore, the total local computing time of CSDs in ABCG and AMND may decrease with $a$. As illustrated in Fig. \ref{fig6}, ABCG and AMND achieve almost the same total local computing time. The reason for this may be that ABCG is an initial input of AMND. In other words, the local computing MDs of AMND may be decided by ABCG.
\begin{figure}[!t]
\centering
\centerline{\includegraphics[width=3.5in,height=3in]{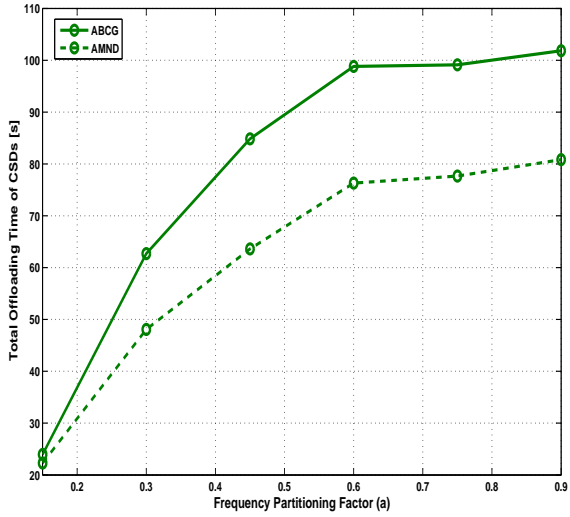}}
\caption{The total offloading time of CSDs versus frequency partitioning factor $a$ under the file request coefficient $\delta $.}
\label{fig7}
\end{figure}
\par
Fig. \ref{fig7} shows the impacts of frequency partitioning factor $a$ on the total offloading time (delay) of CSDs under the file request coefficient $\delta $. As revealed in Fig. \ref{fig6}, the number of local computing MDs may decrease with $a$. That means the number of edge computing MDs may increase with $a$. Since more served MDs may mean that they can obtain fewer resources for edge computing, the total offloading time of CSDs may increase with $a$. 
\begin{figure}[!t]
\centering
\centerline{\includegraphics[width=3.5in,height=3in]{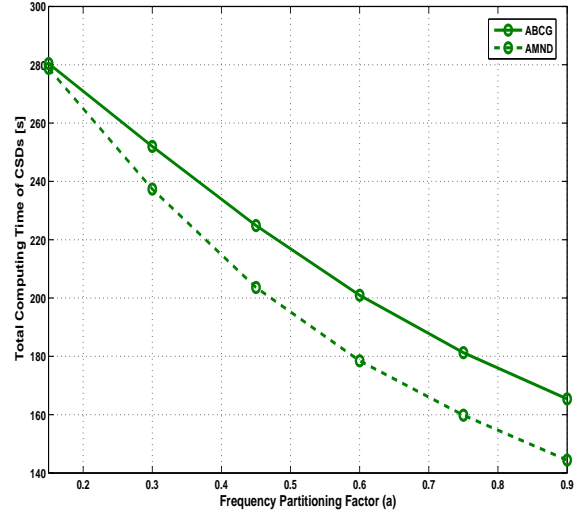}}
\caption{The total time of CSDs versus frequency partitioning factor $a$ under the file request coefficient $\delta $.}
\label{fig8}
\end{figure}
\par
Fig. \ref{fig8} shows the impacts of frequency partitioning factor $a$ on the total time (delay) of CSDs under the file request coefficient $\delta $. According to Figs. \ref{fig6} and \ref{fig7}, we can easily know that the local computing has a major impact on the total computing time of CSDs. Therefore, such time should decrease with $a$.

\begin{figure}[!t]
\centering
\centerline{\includegraphics[width=3.5in,height=3in]{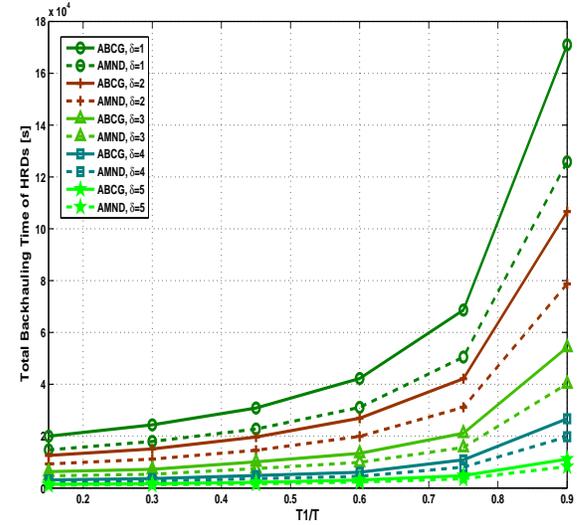}}
\caption{The total (downlink) backhauling time of HRDs versus time partitioning factor ${T_1}/{T}$ under the file request coefficient $\delta $.}
\label{fig9}
\end{figure}
\par
Fig. \ref{fig9} shows the impacts of time partitioning factor ${T_1}/{T}$ on the total (downlink) backhauling time (delay) of HRDs under the file request coefficient $\delta $. Evidently, an increased factor ${T_1}/{T}$ results in fewer and fewer resources utilized by HRDs on downlink backhauling links. Therefore, the total backhauling time of HRDs may increase with such a factor.
\begin{figure}[!t]
\centering
\centerline{\includegraphics[width=3.5in,height=3in]{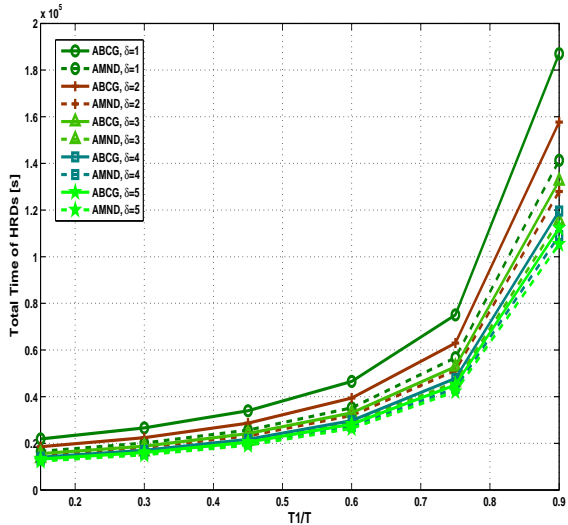}}
\caption{The total time of HRDs versus time partitioning factor ${T_1}/{T}$ under the file request coefficient $\delta $.}
\label{fig10}
\end{figure}
\par
Fig. \ref{fig10} shows the impacts of time partitioning factor ${T_1}/{T}$ on the total time (delay) of HRDs under the file request coefficient $\delta $. As we know, an increased factor ${T_1}/{T}$ results in fewer and fewer resources utilized by HRDs on both downlink access and backhauling links. That means the total time of HRDs may increase with such a factor.
\begin{figure}[!t]
\centering
\centerline{\includegraphics[width=3.5in,height=3in]{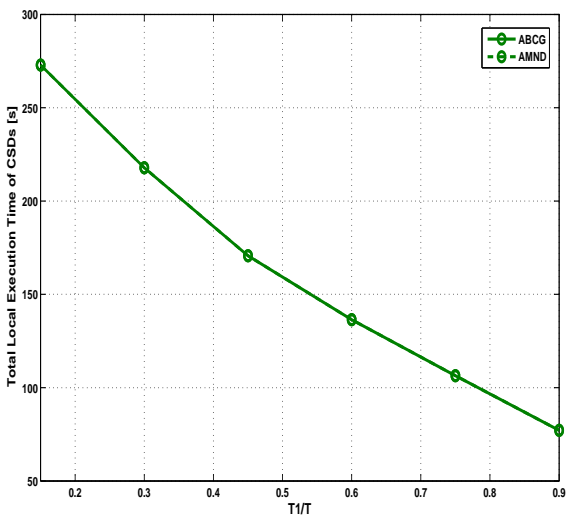}}
\caption{The total local computing time of CSDs versus time partitioning factor ${T_1}/{T}$ under the file request coefficient $\delta $.}
\label{fig11}
\end{figure}
\par
Fig. \ref{fig11} shows the impacts of time partitioning factor ${T_1}/{T}$ on the total local computing (execution) time (delay) of CSDs under the file request coefficient $\delta $. Evidently, an increased factor ${T_1}/{T}$ often means that CSDs have higher uplink data rates, and more and more CSDs may tend to offload their computing tasks to SBSs. Therefore, the total local computing time of CSDs in ABCG and AMND may decrease with such a factor. Similar to Fig. \ref{fig6}, ABCG and AMND achieve almost the same total local computing time.
\begin{figure}[!t]
\centering
\centerline{\includegraphics[width=3.5in,height=3in]{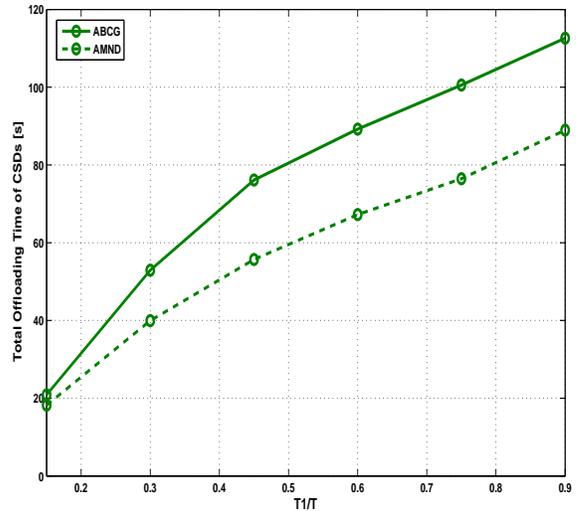}}
\caption{The total offloading time of CSDs versus time partitioning factor ${T_1}/{T}$ under the file request coefficient $\delta $.}
\label{fig12}
\end{figure}
\par
Fig. \ref{fig12} shows the impacts of time partitioning factor ${T_1}/{T}$ on the total offloading time (delay) of CSDs under the file request coefficient $\delta $. As shown in Fig. \ref{fig11}, the number of local computing MDs may decrease with time partitioning factor ${T_1}/{T}$. That means the number of edge computing MDs may increase with such a factor. Since more served MDs may mean they can utilize fewer resources to complete the edge computing, the total offloading time of CSDs may increase with ${T_1}/{T}$.
\begin{figure}[!t]
\centering
\centerline{\includegraphics[width=3.5in,height=3in]{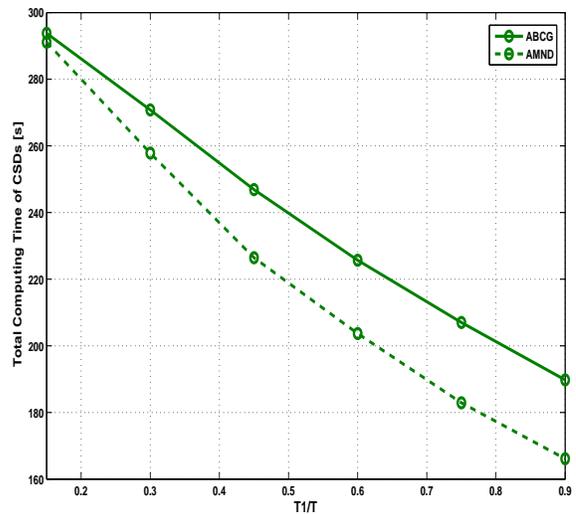}}
\caption{The total computing time of CSDs versus time partitioning factor ${T_1}/{T}$ under the file request coefficient $\delta $.}
\label{fig13}
\end{figure}
\par
Fig. \ref{fig13} shows the impacts of time partitioning factor ${T_1}/{T}$ on the total computing time (delay) of CSDs under the file request coefficient $\delta $. According to Figs. \ref{fig11} and \ref{fig12}, we can easily find that the local computing has a major impact on the total computing time of CSDs. Therefore, such time should decrease with time partitioning factor ${T_1}/{T}$.

\section{CONCLUSION}
In this paper, we jointly perform MD association and resource allocation to minimize the sum of MDs' weighted delays for small-cell IoT networks, and introduce orthogonal time and frequency partition mechanism to mitigate the network interference simultaneously. To solve the formulated problem, Algorithm AMND is designed according to coalitional game and convex optimization theorem. Additionally, Algorithm ABCG is designed to initialize Algorithm AMND and used for comparison. Then, the convergence, stability, parallel implement and computation complexity of them are analyzed. In the simulation, we investigated the impacts of time and frequency partitioning factors on the network delay, and then find that Algorithm AMND may achieve a better performance than Algorithm ABCG. Future work can include the implement of power control/allocation, data compression, full duplex and so on.



\begin{thebibliography}{1}

\bibitem{TTaleb2017} T. Taleb, K. Samdanis, B. Mada, \textit{et al.}, ``On multi-access edge computing: a survey of the emerging 5G network edge cloud architecture and orchestration," \textit{IEEE Commun. Surv. Tut.}, vol. 19, no. 3, pp. 1657-1681, Thirdquarter 2017.

\bibitem{YMao2017} Y. Mao, C. You, J. Zhang, \textit{et al.}, ``A survey on mobile edge computing: the communication perspective," \textit{IEEE Commun. Surv. Tut.}, vol. 19, no. 4, pp. 2322-2358, Fourthquarter 2017.

\bibitem{SWang2017} S. Wang, X. Zhang, Y. Zhang, \textit{et al.}, ``A survey on mobile edge networks: convergence of computing, caching and communications," \textit{IEEE Access}, vol. 5, pp. 6757-6779, 2017.

\bibitem{JYao2019} J. Yao, T. Han, N. Ansari, ``On mobile edge caching," \textit{IEEE Commun. Surv. Tut.}, vol. 21, no. 3, pp. 2525-2553, Thirdquarter 2019.

\bibitem{PMach2017} P. Mach and Z. Becvar, ``Mobile edge computing: a survey on architecture and computation offloading," \textit{IEEE Commun. Surv. Tut.}, vol. 19, no. 3, pp. 1628-1656, Thirdquarter 2017.

\bibitem{FGuo2018Dec} F. Guo, H. Zhang, H. Ji, \textit{et al.}, ``An efficient computation offloading management scheme in the densely deployed small cell networks with mobile edge computing," \textit{IEEE/ACM Trans. Network.}, vol. 26, no. 6, pp. 2651-2664, Dec. 2018.

\bibitem{LChen2018Aug} L. Chen, S. Zhou, J. Xu, ``Computation peer offloading for energy-constrained mobile edge computing in small-cell networks," \textit{IEEE/ACM Trans. Network.}, vol. 26, no. 4, pp. 1619-1632, Aug. 2018.

\bibitem{JZhao2019Aug} J. Zhao, Q. Li, Y. Gong, \textit{et al.}, ``Computation offloading and resource allocation for cloud assisted mobile edge computing in vehicular networks," \textit{IEEE Trans. Veh. Technol.}, vol. 68, no. 8, pp. 7944-7956, Aug. 2019.

\bibitem{KPoularakis2016Apr} K. Poularakis, G. Iosifidis, V. Sourlas\textit{et al.}, ``Exploiting caching and multicast for 5G wireless networks," \textit{IEEE Trans. Wireless Commun.}, vol. 15, no. 4, pp. 2995-3007, Apr. 2016.

\bibitem{LWang2018Sep} L. Wang, K. Wong, S. Lambotharan,\textit{et al.}, ``Edge caching in dense heterogeneous cellular networks with massive MIMO-aided self-backhaul,"  \textit{IEEE Trans. Wireless Commun.}, vol. 17, no. 9, pp. 6360-6372, Sep. 2018.

\bibitem{NAbbas2018Sep} N. Abbas, H. Hajj, S. Sharafeddine,\textit{et al.}, ``Traffic offloading with channel allocation in cache-enabled ultra-dense wireless networks," \textit{IEEE Trans. Veh. Technol.}, vol. 67, no. 9, pp. 8723-8737, Sep. 2018.

\bibitem{Zhou2018Mar} T. Zhou, Z. Liu, J. Zhao, \textit{et al.}, ``Joint user association and power control for load balancing in downlink heterogeneous cellular networks," \textit{IEEE Trans. Veh. Technol.}, vol. 67, no. 3, pp. 2582-2593, Mar. 2018.

\bibitem{TZhou2019May} T. Zhou, J. Zhao, D. Qin, \textit{et al.}, ``Green base station assignment for NOMA-enabled HCNs," \textit{IEEE Access}, vol. 7, pp. 53018-53031, May 2019.

\bibitem{JZhao2019Dec} J. Zhao, S. Ni, L. Yang, \textit{et al.}, ``Multiband cooperation for 5G HetNets: a promising network paradigm," \textit{IEEE Veh. Technol. Mag.}, vol. 14, no. 4, pp. 85-93, Dec. 2019.

 \bibitem{QYe2016May} Q. Ye, O. Y. Bursalioglu, H. C. Papadopoulos, \textit{et al.}, ``User association and interference management in massive MIMO HetNets," \textit{ IEEE Trans. Commun.}, vol. 64, no. 5, pp. 2049-2065, May 2016.

 \bibitem{DLiu2016Sec} D. Liu, L. Wang, Y. Chen, \textit{et al.}, ``User association in 5G networks: a survey and an outlook," \textit{IEEE Commun. Surv. Tut.}, vol. 18, no. 2, pp. 1018-1044, Secondquarter 2016.

 \bibitem{YXu2017Mar} Y. Xu and S. Mao, ``User association in massive MIMO HetNets," \textit{IEEE Syst. J.}, vol. 11, no. 1, pp. 7-19, Mar. 2017.

\bibitem{GRen2017Oct} G. Ren, H. Qu, J. Zhao, \textit{et al.}, ``A distributed user association and resource allocation method in cache-enabled small cell networks," \textit{China Commun.}, vol. 14, no. 10, pp. 95-107, Oct. 2017.

\bibitem{THan2017Oct} T. Han, N. Ansari, \textit{et al.}, ``Network utility aware traffic load balancing in backhaul-constrained cache-enabled small cell networks with hybrid power supplies," \textit{IEEE Trans. Mobile Comput.}, vol. 16, no. 10, pp. 2819-2832, Oct. 2017.

\bibitem{JKwak2019Aug} J. Kwak, L. B. Le, H. Kim, \textit{et al.}, ``Two time-scale edge caching and BS association for power-delay tradeoff in multi-cell networks," \textit{IEEE Trans. Commun.}, vol. 67, no. 8, pp. 5506-5519, Aug. 2019.

\bibitem{FGuo2018Jul} F. Guo, H. Zhang, X. Li, \textit{et al.}, ``Joint optimization of caching and association in energy-harvesting-powered small-cell networks," \textit{IEEE Trans. Veh. Technol.}, vol. 67, no. 7, pp. 6469-6480, Jul. 2018.

\bibitem{KGuo2018Nov} K. Guo, C. Yang, T. Liu, ``Jointly optimizing user association and BS muting for cache-enabled networks with network-coded multicast and reconstructed interference cancelation," \textit{IEEE Trans. Commun.}, vol. 66, no. 11, pp. 5539-5553, Nov. 2018.

\bibitem{WTeng2019Oct} W. Teng, M. Sheng, K. Guo, \textit{et al.}, ``Content placement and user association for delay minimization in small cell networks,"  \textit{IEEE Trans. Veh. Technol.}, vol. 68, no. 10, pp. 10201-10215, Oct. 2019.

\bibitem{YWang2016} Y. Wang, X. Tao, X. Zhang, \textit{et al.}, ``Joint caching placement and user association for minimizing user download delay," \textit{IEEE Access}, vol. 4, pp. 8625-8633, 2016.

\bibitem{WJing2019} W. Jing, X. Wen, Z. Lu, \textit{et al.}, ``User-centric delay-aware joint caching and user association optimization in cache-enabled wireless networks,"  \textit{IEEE Access}, vol. 7, pp. 74961-74972, 2019.

\bibitem{YDai2018Dec} Y. Dai, D. Xu, S. Maharjan, \textit{et al.}, ``Joint computation offloading and user association in multi-task mobile edge computing," \textit{IEEE Trans. Veh. Technol.}, vol. 67, no. 12, pp. 12313-12325, Dec. 2018.

\bibitem{YDu2019} Y. Du, J. Li, L. Shi, \textit{et al.}, ``Two-tier matching game in small cell networks for mobile edge computing," \textit{IEEE Trans. Serv. Comput.}, To be published, 2019.

\bibitem{YQian2019Oct} Y. Qian, F. Wang, J. Li, \textit{et al.}, ``User association and path Planning for UAV-aided mobile edge computing with energy restriction," \textit{IEEE Wireless Commun. Lett.}, vol. 8, no. 5, pp. 1312-1315, Oct. 2019.

\bibitem{JZhou2019} J. Zhou, X. Zhang, W. Wang, ``Joint resource allocation and user association for heterogeneous services in multi-access edge computing networks," \textit{IEEE Access}, vol. 7, pp. 12272-12282, 2019.

\bibitem{ZTan2018Feb} Z. Tan, F. R. Yu, X. Li, \textit{et al.}, ``Virtual resource allocation for heterogeneous services in full duplex-enabled SCNs with mobile edge computing and caching," \textit{IEEE Trans. Veh. Technol.}, vol. 67, no. 2, pp. 1794-1808, Feb. 2018.

\bibitem{LPQian2017Sep} L. P. Qian, Y. Wu, H. Zhou, \textit{et al.}, ``Joint uplink base station association and power control for small cell networks with non-orthogonal multiple access," \textit{IEEE Trans. Wireless Commun.}, vol. 16, no. 9, pp. 5567-5582, Sep. 2017.

\bibitem{NZhao2019Oct} N. Zhao, H. Wu and Y. Chen, ``Coalition game-based computation resource allocation for wireless blockchain networks," \textit{IEEE Internet Things J.}, vol. 6, no. 5, pp. 8507-8518, Oct. 2019.

\bibitem{JHoydis2013} J. Hoydis, K. Hosseini, S. Ten Brink, \textit{et al.}, ``Making smart use of excess antennas: Massive MIMO, small cells, and TDD," \textit{Bell Labs Tech. J.}, vol.18, no.2, pp.5-21, 2013.

\bibitem{LZhang2016Jun} L. Zhang, M. Xiao, G. Wu, \textit{et al.}, ``Efficient scheduling and power aAllocation for D2D-assisted wireless caching networks," \textit{IEEE Trans. Commun.}, vol. 64, no. 6, pp. 2438-2452, Jun. 2016.
\end{thebibliography}

\end{document}